\documentclass[11pt,twoside,a4paper,cmspaper,final,collab]{cms-tdr}
\def\svnVersion{614c116-D}\def\svnDate{2026/02/18}\def\cmsCernNoTag{CERN-EP-2022-103}\def\cmsCernDate{\today}\def\cmsMessage{Published in the Journal of High Energy Physics as \href{http://dx.doi.org/10.1007/JHEP07(2023)161}{\doi{10.1007/JHEP07(2023)161} and \href{https://doi.org/10.1007/JHEP07(2026)155}{\doi{10.1007/JHEP07(2026)155}}.}}
\begin{document}\cmsNoteHeader{EXO-21-010}

\newlength\cmsFigWidth
\ifthenelse{\boolean{cms@external}}{\setlength\cmsFigWidth{0.85\columnwidth}}{\setlength\cmsFigWidth{0.4\textwidth}}
\newcommand{\RunLumi}{138\fbinv}
\newcommand{\jecUncert}{2\%\xspace}
\newcommand{\PAX}{{\HepAntiParticle{X}{}{}}\Xspace}
\newcommand{\PY}{{\HepParticle{Y}{}{}}\Xspace}
\newcommand{\PS}{{\HepParticle{S}{}{}}\Xspace}
\newcommand{\PSuu}{\ensuremath{\PS_{\PQu\PQu}}\xspace}
\newcommand{\PSdd}{\ensuremath{\PS_{\PQd\PQd}}\xspace}
\newcommand{\mjj}{\ensuremath{\overline{m}_{jj}}\xspace}
\newcommand{\mjjjj}{\ensuremath{m_{4j}}\xspace}
\newcommand{\atrue}{\ensuremath{\alpha_{\mathrm{true}}}\xspace}
\newcommand{\yuu}{\ensuremath{y_{uu}}\xspace}
\newcommand{\ydd}{\ensuremath{y_{dd}}\xspace}
\newcommand{\ychi}{\ensuremath{y_{\PGc}}\xspace}
\newcommand{\yomega}{\ensuremath{y_{\PGo}}\xspace}
\renewcommand{\appendixname}{Addendum}

\cmsNoteHeader{EXO-21-010}
\title{Search for resonant and nonresonant production of pairs of dijet resonances in proton-proton collisions at \texorpdfstring{$\sqrt{s} = 13\TeV$}{sqrt(s) = 13 TeV}}

\abstract{A search for pairs of dijet resonances with the same mass is conducted in final states with at least four jets.  Results are presented separately for the case where the four jet production proceeds via an intermediate resonant state and for nonresonant production. The search uses a data sample corresponding to an integrated luminosity of \RunLumi collected by the CMS detector in proton-proton collisions at $\sqrt{s}=13\TeV$.  Model-independent limits, at 95\% confidence level, are reported on the production cross section of  four-jet and dijet resonances.  These first LHC limits on resonant pair production of dijet resonances via high mass intermediate states are applied to a signal model of diquarks  that decay into pairs of vector-like quarks, excluding diquark masses below 7.6\TeV for a  particular model scenario. There are two events in the tails of the distributions, each with a four-jet mass of 8\TeV and an average dijet mass of 2\TeV,  resulting in local and global significances of 3.9 and 1.6 standard deviations, respectively, if interpreted as a signal. The nonresonant search excludes pair production of top squarks with masses between  0.50\TeV to 0.77\TeV, with the exception of a small interval between 0.52 and 0.58\TeV,  for supersymmetric $R$-parity-violating decays to quark pairs, significantly extending previous limits.  Here, the most significant excess above the predicted background occurs at an average dijet mass of 0.95\TeV, for which the local and global significances  are 3.6 and 2.5 standard deviations, respectively.}

\hypersetup{
pdfauthor={CMS Collaboration},
pdftitle={Search for resonant and nonresonant production of pairs of dijet resonances in proton-proton collisions at sqrt(s) = 13 TeV},
pdfsubject={CMS},
pdfkeywords={CMS, paired dijets, resonances}}

\maketitle 

\section{Introduction}
\label{sec:Introduction}

The potential to discover physics beyond the standard model (BSM) is the motivation of searches for dijet resonances. These searches look for a 
pair of jets coming from the decay of a new particle, \PX. 

The simplest searches, which have been conducted at the CERN LHC 
~\cite{Sirunyan:2019vgj,Aad:2019hjw,Sirunyan:2018xlo,Aaboud:2017yvp,Sirunyan:2016iap,Khachatryan:2015dcf,ATLAS:2015nsi,Khachatryan:2016ecr,Khachatryan:2015sja,Aad:2014aqa,Chatrchyan:2013qhXX,CMS:2012yf,Aad201237,ATLAS:2012pu,Chatrchyan2011123,Aad:2011aj,Khachatryan:2010jd,ATLAS2010} 
using strategies reviewed in Ref.~\cite{dijetreview}, are for $s$-channel single production of dijet resonances produced in proton-proton (pp) collisions, where there 
are two jets in the final  state: $\Pp\Pp \to \PX \to jj$. Because of this simple topology, where \PX is produced directly from the  partons in the 
proton and must decay to the same partons, searches for single production place tight constraints on many models of BSM physics. 

\begin{figure}[ht]
\centering
\includegraphics[width=0.45\textwidth]{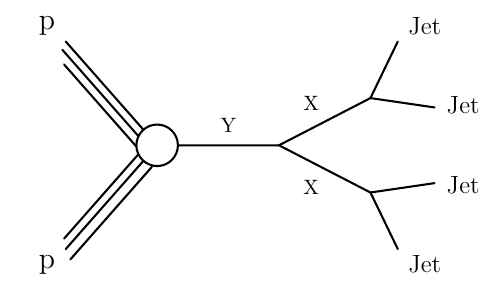}
\hspace{1cm}
\includegraphics[width=0.45\textwidth]{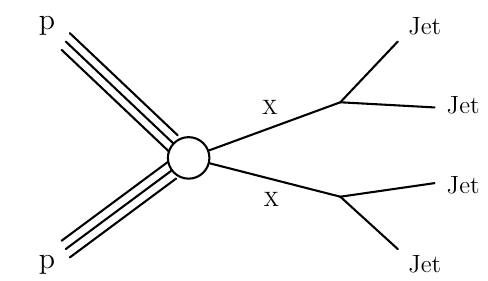}
\caption{(Left) Resonant production via a particle, \PY, of pairs of dijet resonances, \PX. (Right) Nonresonant production of pairs of dijet resonances.} 
\label{fig:feynman} 
\end{figure}

A more complex process is pair production of a dijet resonance $\PX$ decaying to two jets and therefore yielding four jets in the final state.  
Two modes of production of the pair of new  particles are possible, resonant production with an intermediate resonance \PY and non-resonant production, both shown in Fig.~\ref{fig:feynman}, and each mode presents a unique opportunity for discovery. 
We present a generic search for pairs of narrow dijet resonances in each production mode with data corresponding to an integrated 
luminosity of \RunLumi collected in 2016--2018 with the CMS detector at the LHC,  and we use results obtained from each mode to constrain a benchmark 
model, as discussed below. 

Prior searches were for Lorentz-boosted low mass resonance pairs, produced via both resonant~\cite{ATLAS:2020iwa} and nonresonant~\cite{Sirunyan:2018rlj} mechanisms, and have been conducted using jet substructure techniques to identify boosted resonances inside of jets. This analysis considers pairs of resolved dijet resonances, \PX, where both jets within each dijet resonance are individually reconstructed, allowing the resonant and nonresonant searches to be sensitive to high resonance masses: \PX masses greater than 0.5\TeV and \PY masses greater than 2.0\TeV.

First, we consider resonant production of pairs of dijet resonances,
\begin{linenomath} 
\begin{equation} 
   \Pp\Pp \to \PY \to \PX\PX \to (jj)(jj),
   \label{eq:first}
\end{equation}
\end{linenomath}
where the intermediate state is a massive new particle, \PY, decaying to identical dijet resonances, \PX. As a benchmark, we consider a diquark 
model~\cite{Robertpaper}, where the intermediate state is a diquark, $\PS$, produced from the annihilation of two up quarks, and the dijet resonance is a vector-like quark, \PGc,
that decays to an up quark (\PQu) and a gluon (\Pg) only;
\begin{linenomath} 
\begin{equation}
  \PQu\PQu \to \PS \to \PGc\PGc \to (\PQu\Pg)(\PQu\Pg).
  \label{eq:resonant}
\end{equation}
\end{linenomath}
This scalar diquark is a good benchmark because it is produced with a large cross section, due to the high probability of finding up quarks at high 
fractional momentum within the proton. 

This is the first time CMS has searched for resonant pair production of dijet resonances via high mass intermediate states. We are 
further motivated to conduct this search 
by an event, a clear candidate for resonant production of pairs of dijet resonances, that was seen and  presented by a CMS search in the dijet final 
state~\cite{Sirunyan:2019vgj}. As discussed in Ref.~\cite{Robertpaper}, that search was not optimized for this signal and could not quantify its 
significance, motivating a dedicated search in the four-jet final state. Here we present an optimized search for resonant production of pairs of 
dijet resonances, exploring a wide range of four-jet and dijet masses.

Second, we consider nonresonant production of pairs of dijet resonances
\begin{linenomath} 
\begin{equation}
  \Pp\Pp \to \PX\PX \to (jj)(jj).
  \label{eq:nonresonant}
\end{equation}
\end{linenomath}

As a benchmark, we consider the $R$-parity-violating (RPV) supersymmetry (SUSY) model~\cite{Barbier:2004ez}, where the dijet resonance is a top squark, $\PSQt$,
with RPV decays to down (\PQd) and strange (\PQs) quarks
\begin{linenomath} 
\begin{equation}
   \Pp\Pp \to \PSQt\PSQt^{*} \to (\PAQd\PAQs)(\PQd\PQs).
\end{equation}
\end{linenomath}
Nonresonant production, and this benchmark model, have been searched for by the CMS and ATLAS Collaborations at $\sqrt{s}=7$\TeV~\cite{Chatrchyan:2013izb,ATLAS:2012ds}, 8\TeV~\cite{Khachatryan:2014lpa,ATLAS:2016lmi}, and 13\TeV~\cite{Sirunyan:2018rlj,Aaboud:2017nmi}. Other relevant models of both resonant and non-resonant production are discussed in Refs.~\cite{Robertpaper,Bai:2010dj,Kilic:2008ub, Kilic:2008pm, Agashe:2020wph}.

Tabulated results are provided in the HEPData record for this analysis~\cite{HEPData}.

\section{The CMS detector}
\label{sec:detector}

A detailed description of the CMS detector and its coordinate system, including definitions of the azimuthal angle $\phi$ and
pseudorapidity $\eta$, is given in Ref.~\cite{refCMS}.
The central feature of the CMS apparatus is a superconducting
solenoid of 6\unit{m} internal diameter providing an axial magnetic field of 3.8\unit{T}.
Within the solenoid volume are located the silicon pixel and strip tracker, and the barrel and
endcap calorimeters ($\abs{\eta}<3.0$), where these latter detectors consist of a lead tungstate crystal electromagnetic
calorimeter, and a brass and scintillator hadron calorimeter.
An iron and quartz-fiber hadron calorimeter is located in the forward region ($3.0<\abs{\eta}<5.0$),
outside the solenoid volume. The muon detection system covers $\abs{\eta}<2.4$ with up to four layers of gas-ionization detectors installed
outside the solenoid and embedded in the layers of the steel flux-return yoke.

\section{Simulated data samples}
\label{sec:SimDataSamples}

A background sample of quantum chromodynamics (QCD) multijet events is produced, for the optimization of the search and qualitative comparisons with the observed data. The generation 
starts from the leading order (LO) QCD $2 \to 2$ processes of jet production, and includes additional jets from QCD initial- and final-state radiation within 
the parton shower. These background predictions are produced with the \PYTHIA 8.205 program, with the CUETP8M1 tune~\cite{Khachatryan:2015pea,Skands:2014pea}, using the parton distribution function (PDF) set NNPDF2.3LO~\cite{Ball:2012cx}.

The benchmark model for the resonant search is based on the aforementioned model of a diquark decaying to pairs of vector-like quarks, defined in 
Eq.~\eqref{eq:resonant}. Specifically, we consider a scalar diquark resonance $\PS$ which decays to two vector-like quarks $\PGc$, which each then decays to 
a quark and gluon pair: 
$\PS \to \PGc\PGc \to (\PQu\Pg)(\PQu\Pg)$. The \MGvATNLO 2.6.5~\cite{madgraph} generator, with additional code specifying the model~\cite{Robertpaper}, is used  to generate these events and \textsc{MadSpin} is used for the vector-like quark decay to a quark and a gluon. 
Events for diquark masses between 2 and 9\TeV and for vector-like quark masses between 0.2 and 3.8\TeV are generated.

The benchmark model for the nonresonant search is the pair production of top squarks defined in Eq.~\eqref{eq:nonresonant}. 
In the RPV SUSY model~\cite{Evans2012}, the top squarks decay to \PQd and \PQs quarks via a three-generation coupling 
$\lambda''_{312}$ with 100\% branching fraction: $\mathcal{B}(\mathrm{ds})=1$. For this model, we also use \MGvATNLO to generate events with up to one additional jet using the MLM matching procedure~\cite{mlmmatch}.  
Samples with top squark masses between 0.5 and 3.0\TeV are used.

For both benchmark signal simulations, \PYTHIA 8.205~\cite{Sjostrand:2014zea} is used to produce the parton shower and the resulting final-state particles. 
For the diquark simulated samples the CP2 underlying event tune~\cite{Khachatryan:2015pea} is used with the NNPDF3.1LO PDF set~\cite{Ball:2014uwa}, and for the top squark samples the CP5 tune~\cite{m_2020_extraction} is used with the NNPDF3.1NNLO PDF set~\cite{Ball:2014uwa}, a next-to next-to leading order PDF.

The simulation of the CMS detector for all samples is handled by \GEANTfour~\cite{geant4}. All samples include the effects of pileup, additional $\Pp\Pp$ interactions in the same or adjacent bunch crossings. This pileup distribution in simulation is weighted to match the one observed in data.

\section{Event reconstruction and selection}
\label{sec:EventReconstructionAndSelection}

A particle-flow (PF) event algorithm aims to reconstruct and identify each individual particle in an event, with an optimized combination of information
from the various elements of the CMS detector~\cite{CMS-PRF-14-001}. Particles are classified as muons, electrons, photons, charged or neutral hadrons. 
To reconstruct jets from the input particles, the anti-\kt algorithm~\cite{Cacciari:2005hq,Cacciari:2008gp} is used with a distance parameter of 0.4 as implemented in
the \FASTJET package~\cite{Cacciari:2011ma}. The primary vertex (PV) is taken to be the vertex corresponding to the hardest scattering in the event, evaluated using tracking information alone, as described in Section 9.4.1 of Ref.~\cite{CMS-TDR-15-02}.
Charged PF candidates not originating from the primary vertex are removed prior to the jet finding. 

Events are selected using a two-tier trigger system~\cite{Khachatryan:2016bia}. Events satisfying loose jet requirements at the first-level (L1) trigger
are examined by the high-level trigger (HLT) system. At L1, single-jet triggers that require at least one jet in the event to exceed a predefined transverse momentum (\pt) threshold are used.
Triggers that require \HT to exceed a threshold, where \HT is the scalar sum of jet \pt for all jets in the event with $\pt>30$\GeV
and $\abs{\eta}<3.0$, are also used. The HLT requires \HT$>1050$\GeV or at least one jet reconstructed with an anti-\kt distance parameter of 0.8 and $\pt>550$\GeV.

The jet momenta and energies are corrected using calibration factors obtained from simulation, test beam results, and $\Pp\Pp$ collision
data at $\sqrt{s}=13$\TeV. The methods described in Ref.~\cite{Khachatryan:2016kdb} are used and all in-situ calibrations are obtained from the
current data. Jets are required to have $\pt>80$\GeV and $\abs{\eta}<2.5$.  The four jets with the largest \pt are defined as the four leading jets. Jet
identification criteria are applied to remove spurious jets associated with calorimeter noise, as well as those associated with muon and electron
candidates that are either misreconstructed or isolated~\cite{CMS:2017wyc}. For all jets, we require that  the component of 
the jet energy arising from neutral hadrons and the component arising from photons, are each less than 90\% of the total jet 
energy. For jets within the fiducial tracker coverage we additionally require the jet to have nonzero charged-hadron energy,
and electron and muon energies to be less than 90\%  and 80\% of the total jet energy respectively. An event is rejected if 
any of the four  leading jets fails these jet identification criteria. 

The dijet pairs are constructed from the four leading jets. In the following definitions the subscripts 1, and 2 refer to the two dijet systems: dijet 1 and dijet 2. 
There are three possible combinations of four jets making two dijet systems. 
A jet pairing algorithm is used to choose one of these three combinations to form the pair of dijets, and the jets are relabeled with indices $\textit{a, b, c}$ and $\textit{d}$.
The pairing algorithm uses the $\eta$-$\phi$ space separations between the jets in each dijet, $\Delta R_{1}$ and $\Delta R_{2}$, given by
  \begin{linenomath} 
		\begin{equation}		
                  \Delta R_{1}  = \sqrt{(\eta_{j_a} - \eta_{j_b})^2 + (\phi_{j_a} - \phi_{j_b})^2},\quad \Delta R_{2} = \sqrt{(\eta_{j_c} - \eta_{j_d})^2 + (\phi_{j_c} - \phi_{j_d})^2},
		\label{eq:deltaR1}		
                \end{equation}
 \end{linenomath} 
where indices $j_a$ and $j_b$ refer to jets in the first dijet system (subscript $1$), and indices $j_c$ and $j_d$ refer to jets 
in the second dijet system (subscript $2$). We find that the 
optimal pairing algorithm is the one minimizing the following measure of the $\eta$-$\phi$ space separations of the jets in the event:
  \begin{linenomath} 
                \begin{equation}
                   \Delta R=\abs{(\Delta R_{1} - 0.8)} + \abs{(\Delta R_{2} - 0.8)}.
                  \label{eq:deltaR}
                \end{equation}
 \end{linenomath} 
This pairing algorithm was  first used in the search for nonresonant production of pairs of dijet resonances~\cite{Sirunyan:2018rlj}. It is 
motivated by the expectation that the jets from a dijet resonance will be closer together than uncorrelated jets, minimizing 
the separation of the jets, while the offsets of 0.8 in Eq.~\eqref{eq:deltaR} reduce pairings where the jets overlap in $\eta$-$\phi$
space.

Once the jet pairing is complete, events are selected that satisfy:
  \begin{linenomath} 
		\begin{equation}
                        \Delta R_{1,2} < 2.0.
		\end{equation}
 \end{linenomath} 
These requirements reject background from hard multijet processes produced by QCD, which does not  naturally give small separations in $\eta$-$\phi$ space between the jets in each dijet in a 
four-jet event. In addition, we require the $\eta$ separation of the two dijets to satisfy:
  \begin{linenomath} 
                \begin{equation}
                         \Delta \eta = \abs{ \eta_{1} - \eta_{2} } < 1.1,
                         \label{eq:deltaEta}
                \end{equation} 
 \end{linenomath}           
which further suppresses background from QCD $t$-channel production compared to the signal from $s$-channel production. Finally, 
the asymmetry in dijet mass between the two dijets is required to be small,
  \begin{linenomath} 
                \begin{equation}
                      \text{asymmetry} = \frac{\abs{m_{1}-m_{2}}}{m_{1}+m_{2}} < 0.1,
                       \label{eq:deltaM}
                \end{equation}
 \end{linenomath} 
selecting dijet pairs of approximately equal mass, a property of pairs of identical resonances, but not of the QCD background.
Here, $m_1$ and $m_2$ denote the reconstructed mass of each dijet pair. Throughout the paper, the small letter $m$ denotes reconstructed
dijet and four-jet masses, and the capital letter $M$ denotes true resonance masses.

An analysis optimization procedure was performed using Monte Carlo (MC) simulations of signals and background. In addition to the 
optimal pairing method that minimized $\Delta R$ in Eq.~\eqref{eq:deltaR}, we also considered a second method that minimized 
the 
asymmetry, a third utilizing both $\Delta R$ and asymmetry, a fourth minimizing the difference between $\Delta R_{1}$ and 
$\Delta R_{2}$, a fifth minimizing the sum of the average dijet mass and the difference in dijet mass, and finally a sixth 
method minimizing both the average dijet mass and the  difference in dijet mass. We varied the thresholds for the event 
selection requirements in Eqs.~(\ref{eq:deltaR1}--\ref{eq:deltaM}), and determined the expected 
signal significance at both low and high resonance masses, for both resonant and nonresonant signals. The optimal dijet 
pairing method and event selections, presented in Eqs.~(\ref{eq:deltaR}--\ref{eq:deltaM}), are the ones that maximized the 
expected signal significance for the full analysis. The different jet pairings explored do not significantly affect the resonant search, 
whose final observable is the four-jet mass, and give similar results for the non-resonant search.

\section{Analysis strategy}
\label{sec:AnalysisStrategy}

Natural variables for the two searches are derived from the properties of the production processes. The two dijet resonances 
$\PX\PX$ have equal mass, so one-dimensional (1D) distributions of the average dijet mass $\overline{m}_{jj}= (m_1 + m_2)/2$ 
can be used to search for the nonresonant production of this pair. 
Resonant pair  production, $\PY \to \PX\PX$, additionally produces a localized excess in a second variable, the 
four-jet mass $m_{4j}$, clearly visible in the two-dimensional (2D) distribution of $\overline{m}_{jj}$ vs. $m_{4j}$. 

\begin{figure}[ht]
\centering
\includegraphics[width=0.9\textwidth]{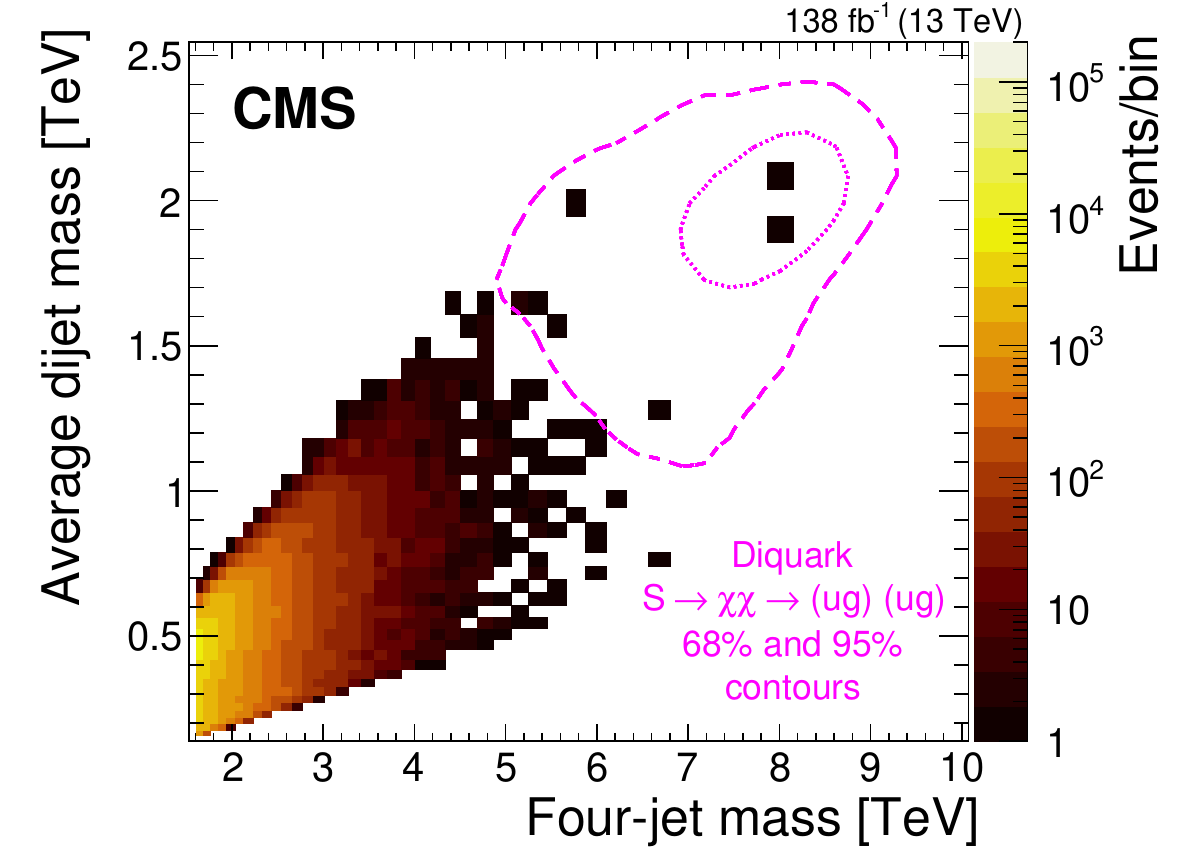}   
\caption{Numbers of events observed (color scale) within bins of the four-jet mass and the average mass of the two dijets. The dotted and dashed curves show the 68\% and 95\% probability contours, respectively, from a signal simulation of a diquark with a mass of 8.4\TeV, decaying to a pair of vector-like quarks, each
with a mass of 2.1\TeV.}
\label{fig:M4J_vs_alpha}
\end{figure}

Figure~\ref{fig:M4J_vs_alpha} shows the 2D distribution of the observed data for $m_{4j}$ vs. $\overline{m}_{jj}$. Near $m_{4j}\sim 8\TeV$ and $\overline{m}_{jj}\sim 2\TeV$ we observe two events, candidates for a resonant  signal. For comparison,  Fig.~\ref{fig:M4J_vs_alpha} also shows the 68 and 95\% probability contours of the signal, respectively, 
from a simulation of our benchmark model of resonant pair production, for a diquark of mass 8.4\TeV and a vector-like quark mass of 2.1\TeV. 
The challenge in evaluating the significance of this signal, or any 
signal, is to understand the background. Perturbative QCD predictions of multijet production have too many theoretical uncertainties to reliably model this 
background. We instead use an approach that does not rely on simulation, by fitting a smooth background parameterization function to the observed data. We do this with a set of 1D 
distributions that span the 2D space.

The seemingly natural approach of binning the search would be to construct the set of 1D distributions of the observed number of events as a function of $m_{4j}$, in bins of $\overline{m}_{jj}$, or vice versa. However, Fig.~\ref{fig:M4J_vs_alpha} shows that due to correlations between those two variables this approach would introduce sculpting of the distributions, which would limit the range of new particle masses we could search for. Hence, we define a new dimensionless variable, $\alpha = \overline{m}_{jj}/m_{4j}$, which is the ratio of the two defined variables of resonant production, and a measure of the boost of the two dijet systems for both production modes. For fixed values of $\alpha$, the number of events observed decreases smoothly with increasing $m_{4j}$, or increasing $\overline{m}_{jj}$. Therefore, binning in a $\alpha$ provides a wide unbiased range of $m_{4j}$ or $\overline{m}_{jj}$ because the distributions are not sculpted or truncated by the boundaries. Thus $\alpha$ is a good variable to bin the 2D data to construct a set of unbiased 1D distributions for the data-driven search.

\begin{figure}[ht]
\centering
\includegraphics[width=0.45\textwidth]{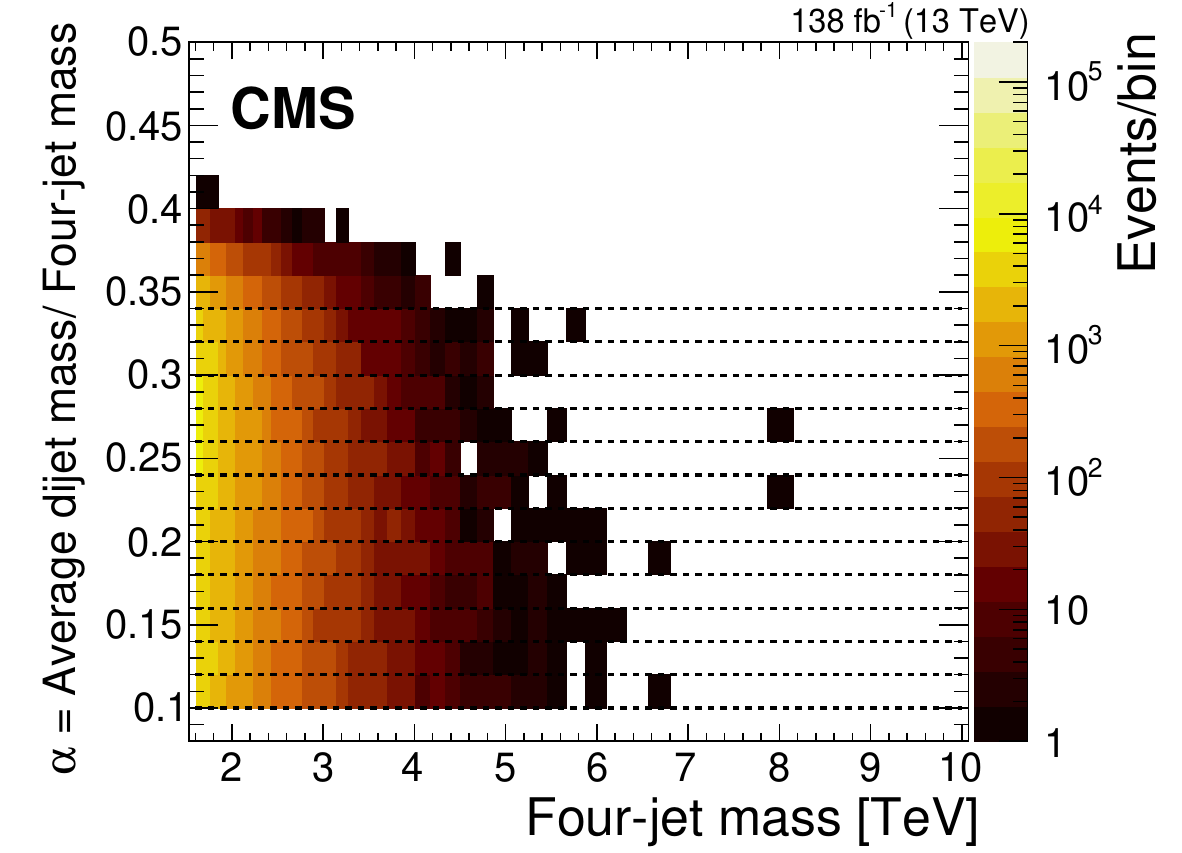}
\hspace{1cm}
\includegraphics[width=0.45\textwidth]{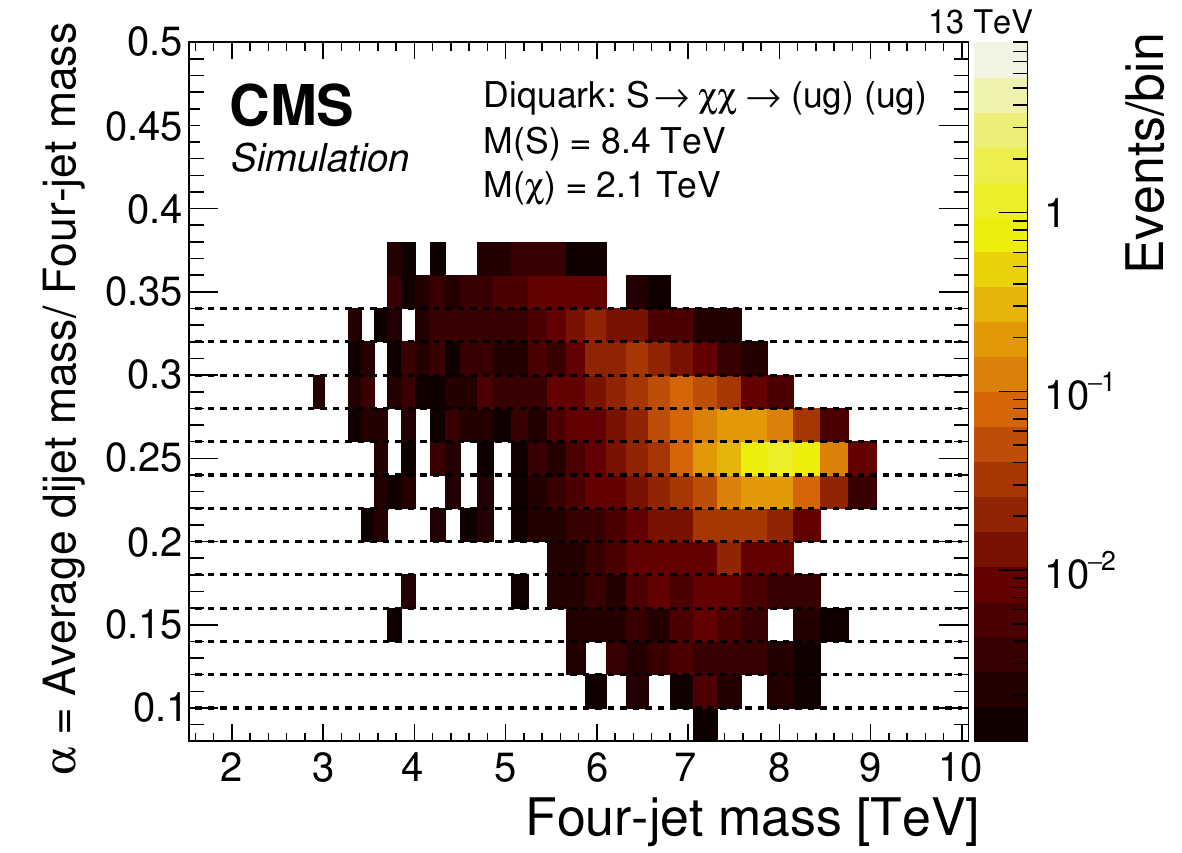}
\caption{(Left) Numbers of events observed (color scale) within bins of the four-jet mass and the ratio $\alpha$, which is the average mass of the two 
dijets divided by the four-jet mass. (Right) Number of events predicted in the same bins by a simulation of a diquark with a mass of 8.4\TeV, 
decaying  to a pair of vector-like quarks, each with a mass of 2.1\TeV. Both plots also show the thirteen $\alpha$ bins used to define the 
four-jet mass distributions used in the resonant search (dashed lines).} 
\label{fig:M4J_vs_alpha_res} 
\end{figure}

\begin{figure}[ht]
\centering
\includegraphics[width=0.45\textwidth]{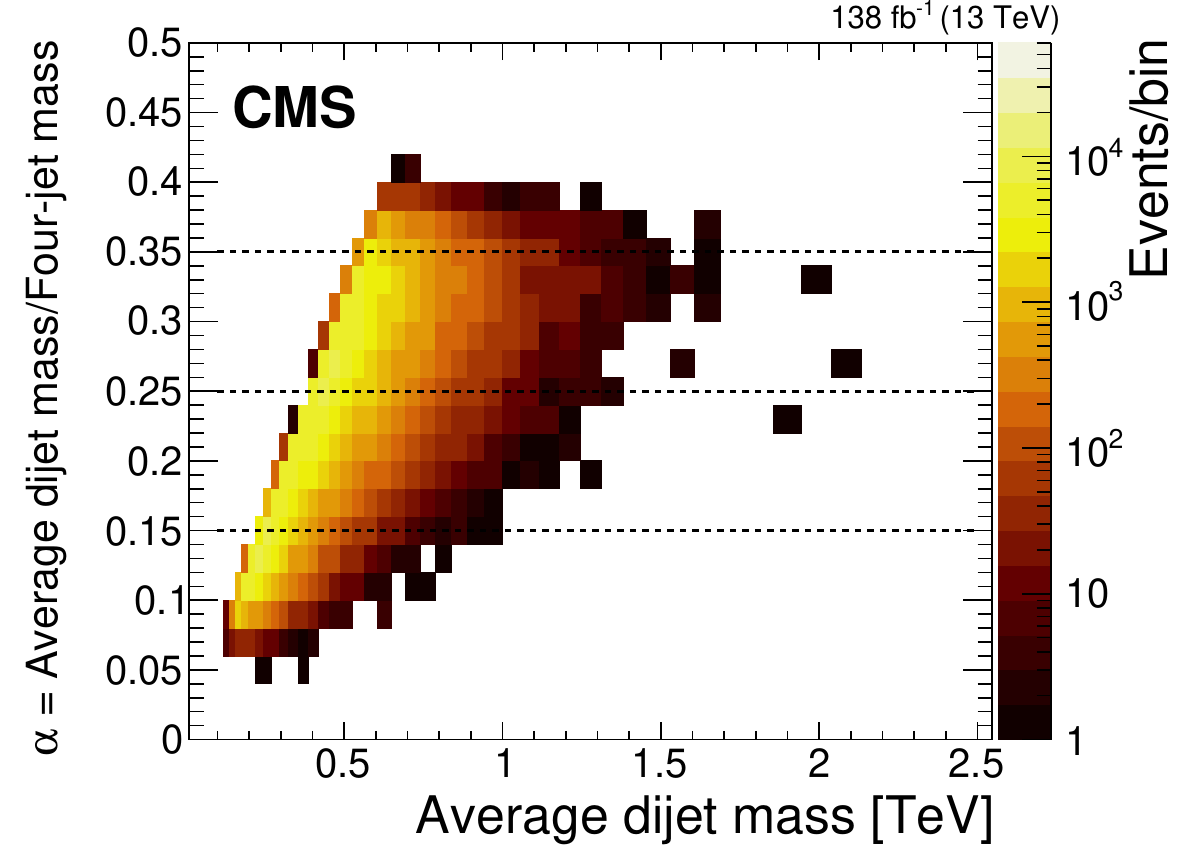}
\hspace{1cm}
\includegraphics[width=0.45\textwidth]{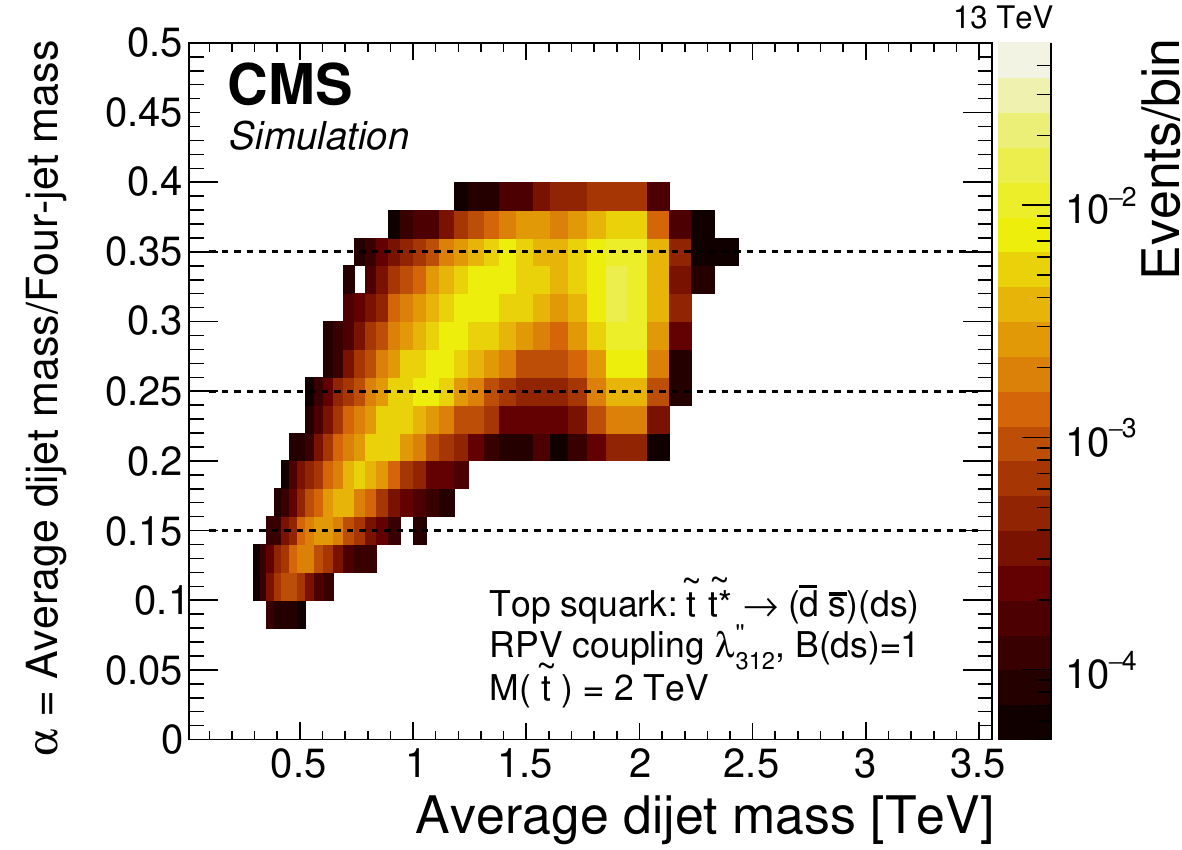}
\caption{(Left) Numbers of events observed (color scale) within bins of the average mass of the two dijets and the ratio $\alpha$ of that mass to the four-jet 
mass. (Right) Number of events predicted in the same bins by a simulation of the production and $R$-parity violating decay of a pair of top squarks with a 
mass of 2\TeV. Both plots also show the three $\alpha$ bins used to define the average dijet mass distributions in the nonresonant search 
(dashed lines).} 
\label{fig:M4J_vs_alpha_non_res} 
\end{figure}

Figures~\ref{fig:M4J_vs_alpha_res} and~\ref{fig:M4J_vs_alpha_non_res} show the 2D distributions of
the number of events, both observed in data and 
expected from a simulated signal, as a function of both $\alpha$ and the  mass variable to be used for the 1D distributions of the search: $m_{4j}$ for the resonant 
search, and $\overline{m}_{jj}$ for the nonresonant search, respectively. The observed 
data for the resonant search in Fig.~\ref{fig:M4J_vs_alpha_res} is distributed with an approximately uniform density in $\alpha$, and for each value of  $\alpha$ there is a wide 
unbiased range of $m_{4j}$ to facilitate estimation of the background.

In addition, although the dijet pairs from a nonresonant signal  are not  produced at a fixed value of $\alpha$, using a binning in 
the nonresonant search exploits the signal distribution in $\alpha$. As can be seen in Fig.~\ref{fig:M4J_vs_alpha_non_res}, properly reconstructed signals (i.e., pairs of dijets with average masses 
near the generated \PSQt mass) tend  to be produced with high $\alpha$. This is especially  true for higher-mass \PSQt pairs, and it is because
the lower values of $\alpha$ only occur when the individual $\PSQt$s are produced with high \pt.  
Hence, examining the four-jet and 
average dijet mass distributions in bins of $\alpha$ is expected to significantly increase the sensitivity of both searches.

The average dijet mass and the four-jet mass are the variables used for the signal extraction in the nonresonant and resonant
search respectively, as described in more detail below. We then optimize, for maximal signal significance, both the number of 
$\alpha$ bins 
considered and the final  observable used to search  for localized excesses. 
The resonant search consists of thirteen simultaneous 1D searches using the $m_{4j}$  distributions within the thirteen $\alpha$ bins shown in Fig.~\ref{fig:M4J_vs_alpha_res}, 
and similarly the nonresonant search uses the  $\overline{m}_{jj}$ distributions in the three $\alpha$ bins in Fig.~\ref{fig:M4J_vs_alpha_non_res}.
For the  resonant search the  lowest $\alpha$ value we consider is 0.1 with a bin width of 0.02, which is comparable to the  reconstructed $\alpha$  resolution, up to an 
$\alpha$ of 0.34, and then we utilize an 
overflow bin for all higher  $\alpha$ values. For the nonresonant search, the optimization yielded a  
bin width of 0.1, for two bins between $\alpha$ of 0.15 and 0.35, and again a final bin is used for higher $\alpha$ values.

Lastly, the searches are performed above mass thresholds for which the trigger is found to be fully efficient.  In
the case of the resonant analysis, the search is performed for
four-jet masses $m_{4j}>1.6$\TeV for all $\alpha$ values.  In the case
of the nonresonant analysis, the thresholds in average dijet mass vary
with $\alpha$: $\overline{m}_{jj}$ is required to be greater than 0.35, 0.53, and 0.61\TeV for the three bins with lower $\alpha$ edges of 0.15, 0.25, and
0.35, respectively. In Fig.~\ref{fig:M4J_vs_alpha_non_res}, this requirement on $\overline{m}_{jj}$ is removed for illustrative purposes, 
and the same events as in Fig.~\ref{fig:M4J_vs_alpha_res} are shown as a function of 
$\alpha$ and $\overline{m}_{jj}$.

\section{Signal simulations}
\label{sec:Signal}

\begin{figure}[ht]
\centering
\includegraphics[width=0.45\textwidth]{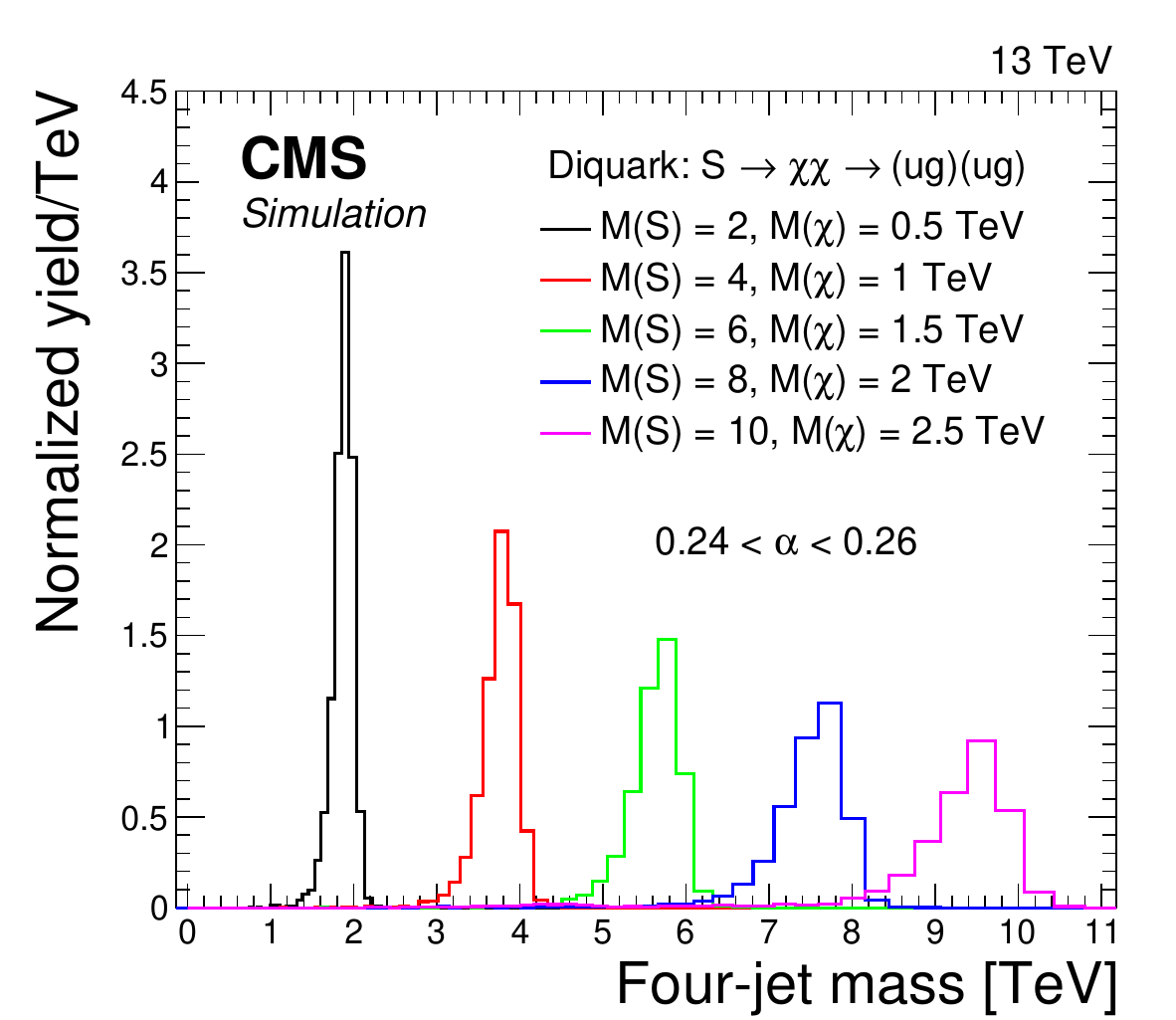}
\includegraphics[width=0.45\textwidth]{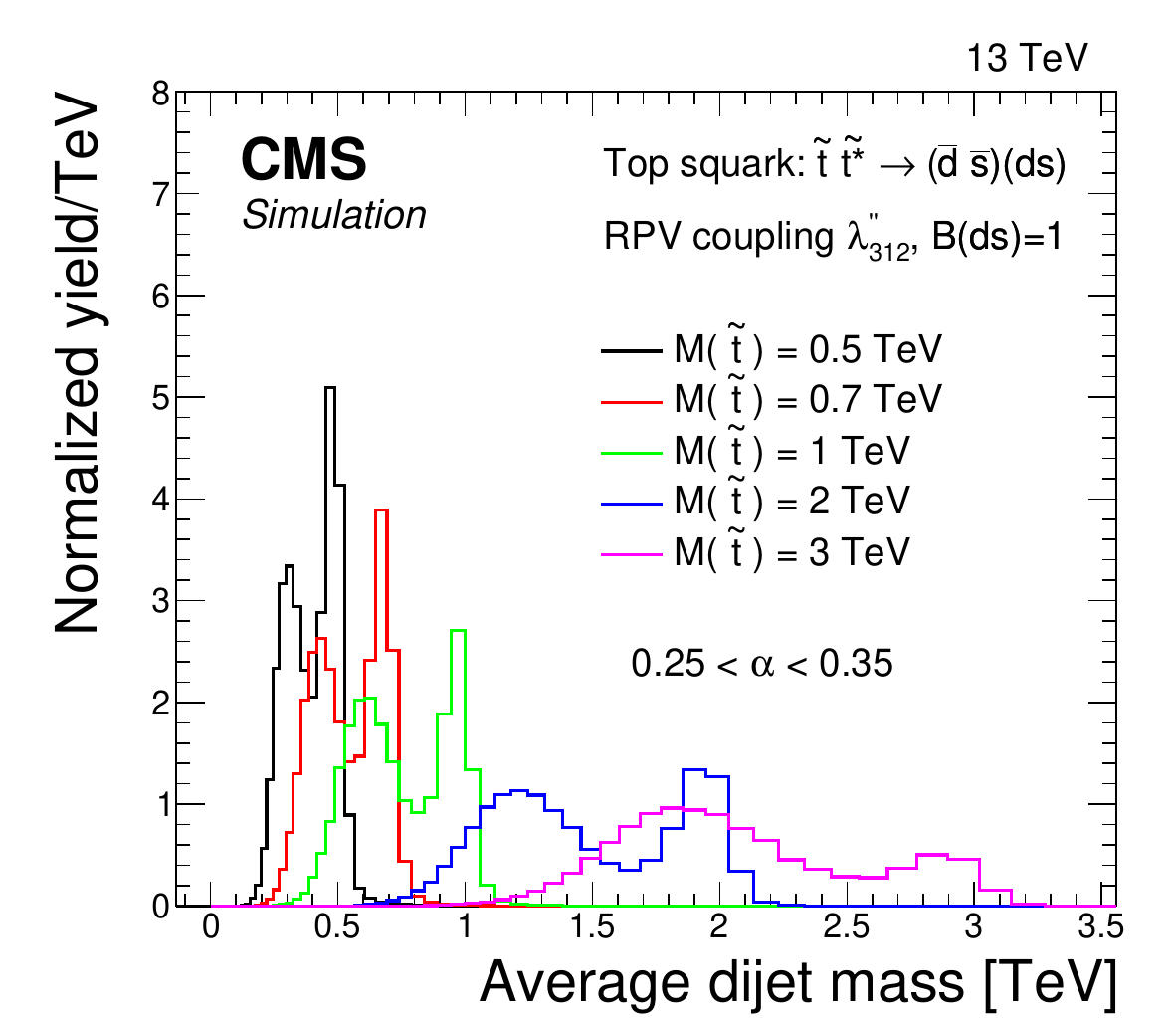}
\caption{Signal differential distributions for the resonant search (left) as functions of four-jet mass, and the nonresonant search (right) 
as functions of the average mass of the two dijets, for various resonance masses, and for the $\alpha$  bins with the largest signal yield in each search. The integral of each distribution has been normalized to unity.}
\label{fig:Shapes_alpha0p25_slices}
\end{figure}

Figure~\ref{fig:Shapes_alpha0p25_slices} shows the shapes of signals in the resonant and nonresonant
searches for various \PS and top squark masses, respectively, for the $\alpha$ bin with the largest  signal present. The distributions of four-jet mass, for signals in 
the resonant search, peak just below 
the input generator mass and have a long tail to low masses from radiation. The distributions  of average dijet mass, for dijet 
resonances in the nonresonant search, also have a 
peak just below the input generator mass from the correct pairings of the four jets, and an additional peak at lower masses from the 
incorrect pairings.

We estimate the pairing efficiency by measuring the fraction of events in which the average reconstructed dijet mass agrees with the true dijet resonance mass within a window of $\pm$3 sigma, where sigma is the average experimental resolution of the reconstructed dijet mass.
For diquarks in the resonant search with $\alpha_{\text{true}}$ = 0.25, shown in Fig.~\ref{fig:Shapes_alpha0p25_slices}, the pairing efficiency is 1.0 for the single $\alpha$ bin $0.24 < \alpha <0.26$. For all $\alpha$ bins combined, the pairing efficiency remains 1.0 at low diquark masses and decreases only a little to 0.9 at the highest diquark masses. For top squarks in the nonresonant search with $0.25 < \alpha <0.35$, also shown in Fig.~\ref{fig:Shapes_alpha0p25_slices}, the pairing efficiency is significantly lower, ranging from 0.4 at the highest top squark mass to 1.0 at the lowest mass.  The pairing efficiency for the resonant search is significantly higher than for the nonresonant search, because the diquark mass kinematically constrains the possible values of average dijet mass, preventing relatively low values. The pairing is more likely to fail for these relatively low values of average dijet mass, producing a secondary peak at lower values of average dijet mass in the nonresonant search, as seen in Fig.~\ref{fig:Shapes_alpha0p25_slices}.

\begin{figure}[ht]
\centering
\includegraphics[width=0.45\textwidth]{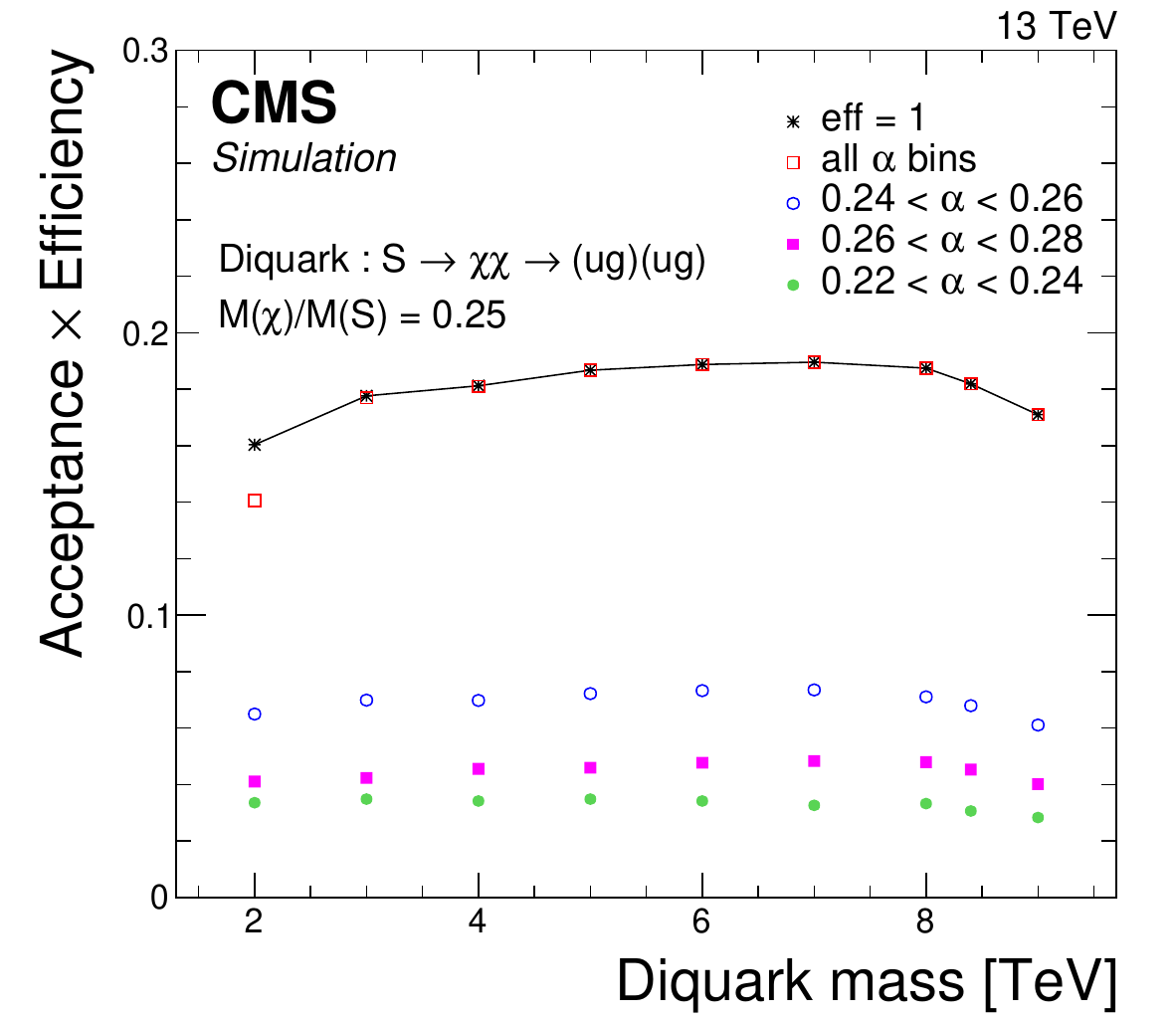}
\includegraphics[width=0.45\textwidth]{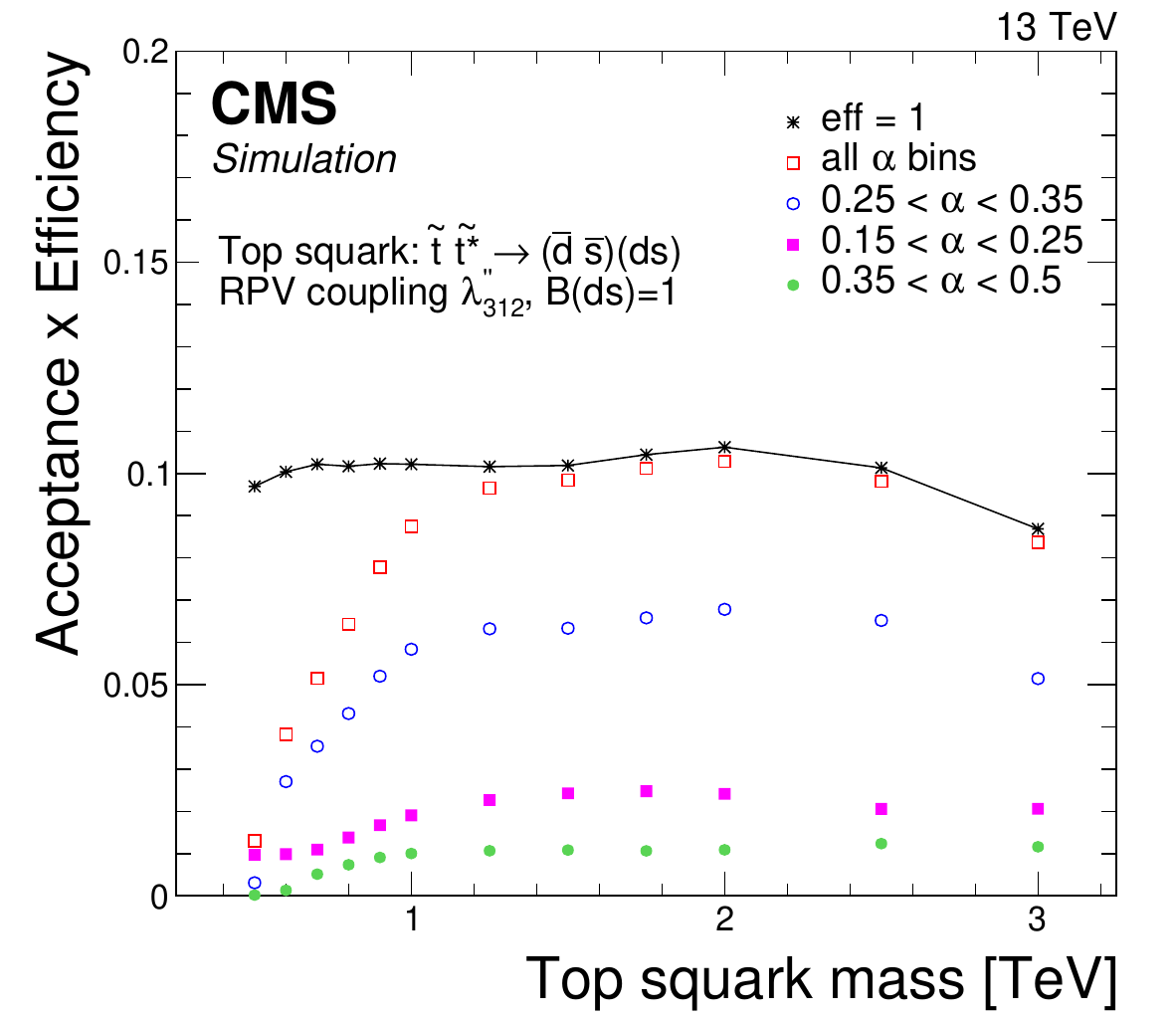}
\caption{The products of acceptance and efficiency of a resonant signal vs. the diquark mass (left), and a nonresonant 
signal vs. the top squark mass (right), inclusively, i.e., for all $\alpha$ values, and for the three $\alpha$ bins that 
contain the majority ($\geq$85\%) of the signal. The case when the efficiency of the mass selection is unity is also shown as a solid line.}
\label{fig:acc_SuuMass}
\end{figure}

The products of acceptance and efficiency, for signals in the resonant and nonresonant searches are shown in Fig.~\ref{fig:acc_SuuMass} as functions of diquark and top 
squark mass. The acceptance is defined as the fraction of events passing the kinematic selection criteria. 
The efficiency is the fraction of signal events satisfying the mass threshold previously discussed: four-jet mass requirement for the resonant 
search, dijet mass selection for the nonresonant 
search. Figure~\ref{fig:acc_SuuMass} also shows the signal acceptance alone, the curve where the efficiency is equal to 1. 
Cross section upper limits in this paper are quoted within this acceptance and corrected for efficiency.
The highest signal acceptance for the resonant search, with $\alpha_{\mathrm{true}} = M(\PX)/M(\PY) = 0.25 $, is naturally for the corresponding bin with $0.24 < \alpha <0.26$. For 
the nonresonant  search, which does not have an $\alpha_{\mathrm{true}}$ or a natural value of $\alpha$, the highest signal acceptance is for the bin with  $0.25 < \alpha 
<0.35$. The acceptance, which originates from the dijet matching and angular separation requirements of Eqs.~(\ref{eq:deltaR1}--\ref{eq:deltaM}), is approximately 
independent of the resonance mass in both searches. However, there is a small decrease in the acceptance at the very highest values of signal mass in Fig.~\ref{fig:acc_SuuMass} for both searches, which is caused by the mass asymmetry requirement. That requirement is more likely to reject events with incorrect dijet pairings, which are more frequent at high signal masses, as previously discussed.

\section {Background estimation}
\label{sec:BackgroundPrediction}

 The background in this search is derived exclusively from data. As in previous resonance 
searches \cite{Sirunyan:2018xlo,Sirunyan:2016iap,Khachatryan:2016ecr,Khachatryan:2015dcf,Khachatryan:2010jd,Chatrchyan2011123,CMS:2012yf,Chatrchyan:2013qhXX,Khachatryan:2015sja,
ATLAS:2015nsi,ATLAS2010,Aad:2011aj,Aad201237,ATLAS:2012pu,Aad:2014aqa,refCDFrun2}, we fit empirical 
functional forms to the observed multijet mass distributions. Our primary background function, which gives the best fit to the differential cross section data without including any signal 
component, is the modified dijet function (ModDijet-3p) given by $\rd\sigma/\rd{m}= p_0(1-x^{1/3})^{p_1}/x^{p_2}~$, where $p_0$, $p_1$, 
$p_2$ are free parameters. Here, $x=m/\sqrt{s}$, with $m$ being the mass used 
for the search, which is the four-jet mass in the resonant search and the  average dijet mass in the nonresonant search. We use two alternate background functions to estimate the  systematic 
uncertainty in the background from variations in the functional form: the dijet function (Dijet-3p) given by $\rd\sigma/\rd{m}= p_0(1-x)^{p_1}/x^{p_2}~$, and the power-law times exponential function (PowExp-3p) given by $\rd\sigma/\rd{m}=p_0\re^{-p_1 x}/x^{p_2}~$.

\begin{figure}[htbp]
\centering
\includegraphics[width=1.\textwidth]{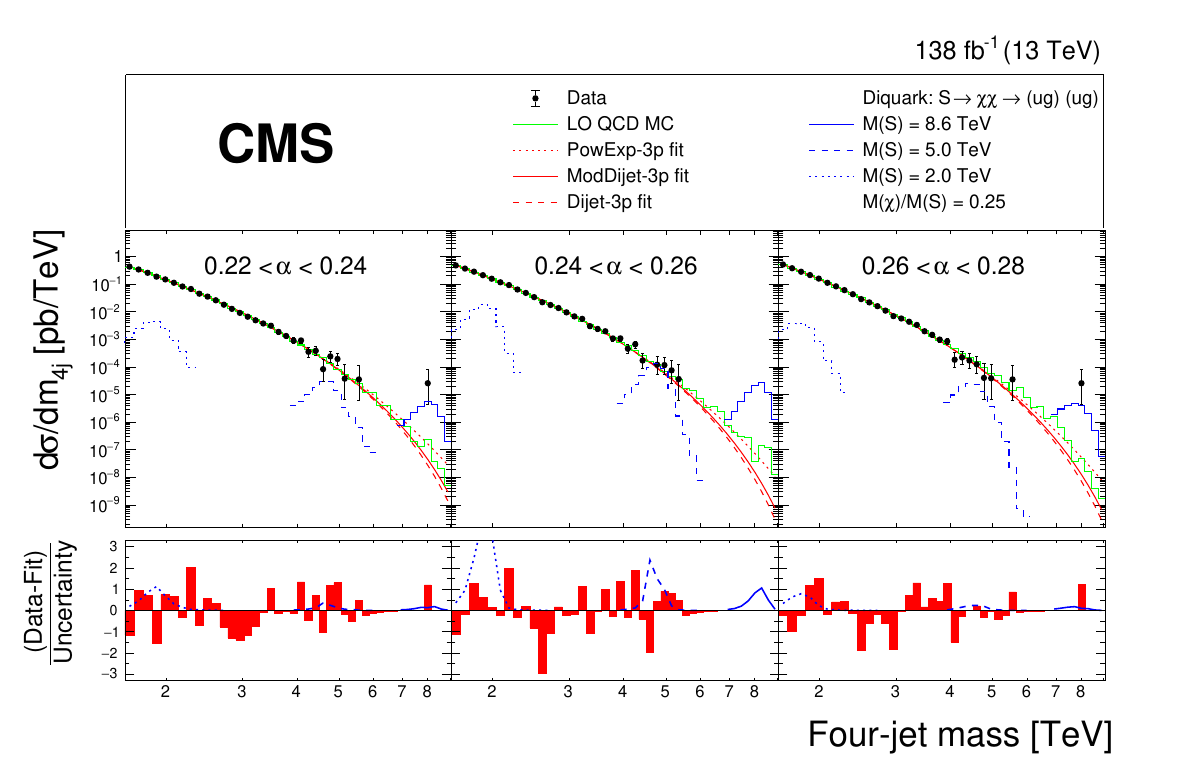}
\caption{
The four-jet mass distributions of the data (points), within three of the thirteen $\alpha$ bins of the resonant search, compared with the simulated LO QCD background distribution  (green histogram) and fitted with three functions: a power-law times an exponential (red dotted), the dijet function (red dashed), and the modified dijet function (red solid), each
function with three free parameters.
Examples of predicted diquark signals are shown, with cross sections equal to the observed upper limits at 95\%  confidence level, for resonance
masses of 2.0 (blue dotted), 5.0 (blue dashed), and 8.6\TeV (blue solid). 
The lower panels show the pulls from the fit of the 
modified dijet function to the data, calculated using the statistical uncertainty of the data.}
\label{fig_dijet2}
\end{figure}

\begin{figure}[ht]
\centering
\includegraphics[width=0.95\textwidth]{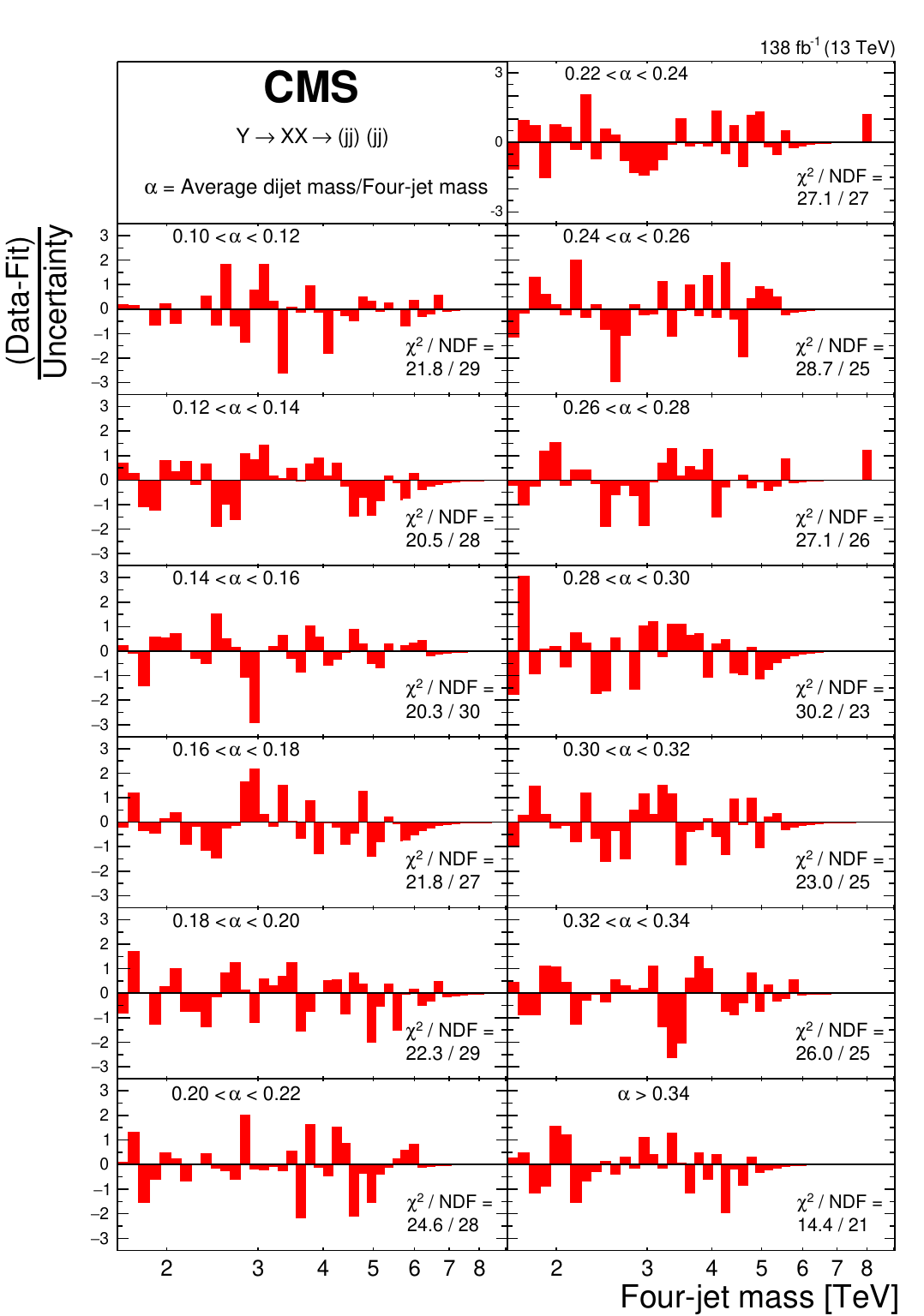}
\caption{The pulls from the fit of the modified dijet function to the four-jet mass distribution, and the reduced chi-squared of the fit 
$\chi^2$/NDF, for all thirteen $\alpha$ bins of the resonant search. 
See Fig.~\ref{fig_dijet2} for the distributions and fits for the $0.22 < \alpha < 0.24$, $0.24 < \alpha < 0.26$ and $0.26 < \alpha < 0.28$ bins.
The pulls are calculated using the statistical uncertainty of the data.}
\label{fig_dijet2a}
\end{figure}

Figure~\ref{fig_dijet2} shows the three functional forms, fitted to examples of the observed differential cross section, as functions of four-jet mass, and the simulation of the QCD 
background, for three out of the thirteen $\alpha$ bins in the resonant search. All functions give a good fit to the data, and a similar prediction of the background to the LO QCD simulation, and 
none of these background estimates appear to account for the two events observed at a four-jet mass of 8\TeV. The three $\alpha$ bins shown in Fig.~\ref{fig_dijet2} contain 85\% of the signal 
for a four-jet resonance, Y, decaying to a pair of dijet  resonances, \PX, with  $\alpha_{\mathrm{true}} = M(\PX)/M(\PY) = 0.25$, for which  
representative signal shapes are also shown in Fig.~\ref{fig_dijet2}. Figure~\ref{fig_dijet2a} shows the pulls from the fit  with the ModDijet-3p function for all thirteen $\alpha$ bins, along with the $\chi^2$ and number of degrees of freedom (NDF) of  each individual fit. 
All the fits adequately describe the data, with individual $p$-value~\cite{GrossVitells} ranging from 0.14 to 0.91. The combined $p$-value is 0.18  for a simultaneous fit of the data with the 
background function, including all thirteen $\alpha$ bins.  This came from conducting a goodness-of-fit test on pseudo-experiments, utilizing a binned likelihood ratio with respect to the saturated model as the test statistic. This probability, and the pulls of the individual fits, indicate that the background model gives a good description of the data overall. 

\begin{figure}[htbp]
\centering
\includegraphics[width=1.\textwidth]{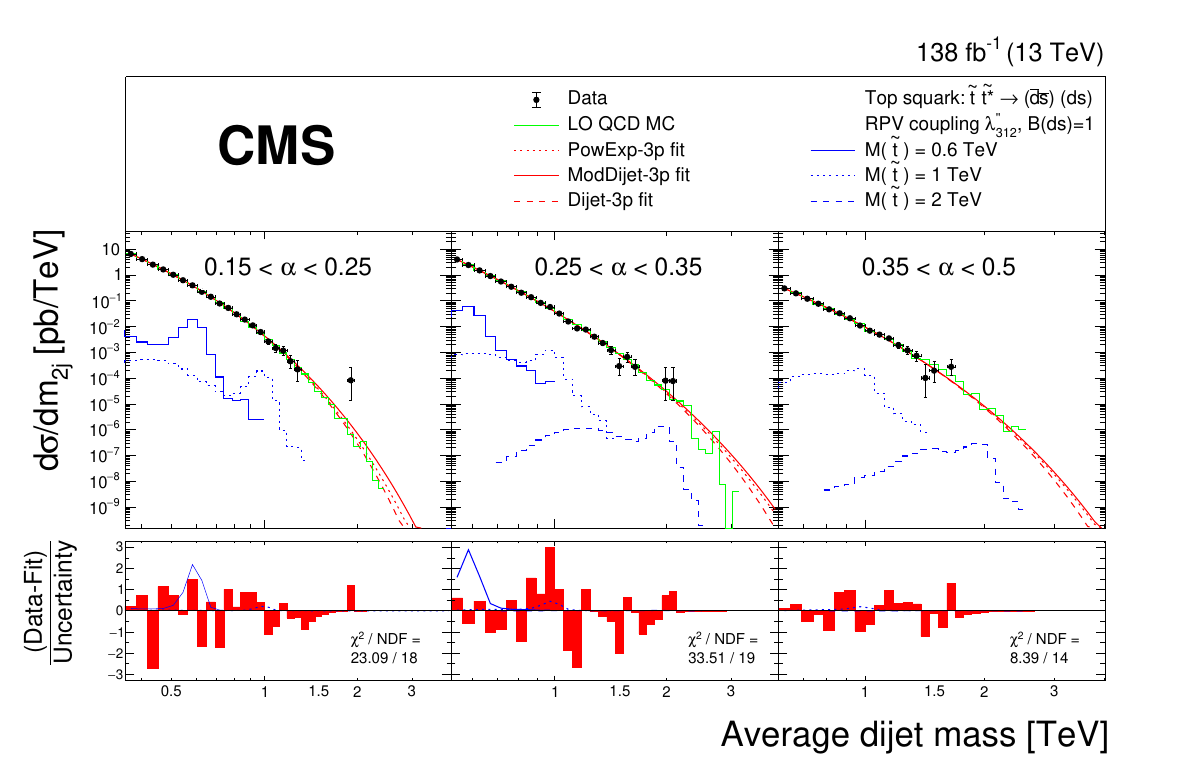}
\caption{The average dijet mass distributions for the data (points), within the three $\alpha$ bins of the nonresonant search, compared with LO 
QCD (green histogram) and fitted with three functions: a power-law times an exponential (red dotted), the dijet function (red dashed), and the modified dijet function (red solid), each
function with three free parameters. Examples of predicted top squark pair production signals are shown, with cross sections equal to the expected SUSY cross sections, for top squark masses of 0.6 (blue solid), 1.0 (blue dotted), and 2.0\TeV (blue dashed). The lower panels show the pulls from the fit of the 
modified dijet function to the dijet mass distribution, calculated using the statistical uncertainty of the data, the aforementioned signals, and the goodness of fit measure 
$\chi^2$/NDF, for all three $\alpha$ bins of the nonresonant search.}
\label{fig:nonresslices}
\end{figure}

For the nonresonant search, Fig.~\ref{fig:nonresslices} shows the data, fits, QCD background simulation, and signal shapes for all three $\alpha$ bins.
The $p$-values of the background fits for the individual $\alpha$ bins range from 0.02 to 0.87, and the combined $p$-value, calculated with the same statistical procedure as in the resonant search, is 0.04. Here the low $p$-values originate from the large absolute values of 
the pulls, near an average dijet mass of roughly 1\TeV for  $0.25 < \alpha <0.35$ in Fig.~\ref{fig:nonresslices}. Similar pulls can also be seen in the corresponding $\alpha$ bins of 
Fig.~\ref{fig_dijet2a}, near a four-jet mass of roughly 3--4\TeV. Only the lowest $\alpha$ bin in the nonresonant search extends to sufficiently small values of average dijet mass to contain a 
noticeable peak for a 0.6\TeV top squark, as shown in Fig.~\ref{fig:nonresslices}. It is the only $\alpha$ bin sensitive to top squark masses between 0.5 and 0.6\TeV.

\clearpage

\section{Results}
\label{sec:Limits}

The search uses a fit of the background function plus the simulated signal shape to the data, taking into account statistical and systematic 
uncertainties. As discussed, the data are the four-jet mass distributions in thirteen bins of $\alpha$ for the resonant search, and the 
average dijet mass distributions in three bins of 
$\alpha$ for the nonresonant search. The fit is done simultaneously for all the bins of  $\alpha$ within each search. The results of the search are the expected and observed upper limits on the 
signal cross section for all experimentally accessible values of resonance mass. For some values of resonance mass, the best fit cross section and observed statistical significance of the signal 
hypothesis are also of interest.

The dominant sources of systematic uncertainty in the signal modeling are the jet energy scale, the jet energy resolution, and the integrated luminosity. 
The uncertainties with the largest impact on the extracted signal yield are the ones relating to background, with the signal modeling related ones being negligible.
The uncertainty in the jet energy scale is below \jecUncert for all values of the four-jet and dijet masses.
This uncertainty is propagated to the limits by shifting the four-jet and dijet mass shapes for the signals by $\pm$\jecUncert.
The uncertainty in the jet energy resolution translates into an uncertainty of 10\% in the resolution of the four-jet and dijet masses~\cite{Khachatryan:2016kdb}, and is propagated
to the limits by observing the effect of increasing the reconstructed width of the four-jet and dijet mass shapes for the signals by 10\%.
The total integrated luminosity has an uncertainty of 1.6\%, the improvement in precision relative to Refs.~\cite{CMS-LUM-17-003,CMS-PAS-LUM-17-004,CMS-PAS-LUM-18-002} 
reflects the uncorrelated time evolution of some systematic effects, and
is propagated to the  normalization of the signal. Changes in the background function used, and the values of the background function parameters as estimated from the fit, 
can introduce changes in the signal yield. All these systematic uncertainties are included in the extracted cross section, limits and significance, as discussed in the next paragraph.

For the signal extraction, we use a multibin counting-experiment likelihood, which is a product of Poisson distributions corresponding to different bins.
We evaluate the likelihood independently at each value of resonance mass from 1.6 to 10\TeV in 0.1\TeV steps for the resonant 
search and, for the nonresonant search, 
in 0.025\TeV steps between 0.5 and 1.0\TeV, 0.05\TeV steps between 1.0 and 1.8\TeV, and 0.2\TeV steps between 1.8 and 3.0\TeV.
These step sizes are comparable to the experimental mass resolution at the lower 
edge of each resonance mass interval.
The sources of systematic uncertainty are implemented as nuisance parameters in the likelihood model, with Gaussian constraints for the
jet energy scale and resolution, and log-normal constraints for the integrated luminosity. 
The main source of the uncertainty in the background modeling comes from the choice of the background function, and the range of possible values of the parameters of the background function.
The parameters of the empirical functional form used to describe
the background are considered as freely floating nuisances, and are evaluated via profiling. The discrete profiling method \cite{Dauncey:2014xga} is used for considering the choice of the functional form as a discrete nuisance parameter,  which is profiled in an
analogous way to continuous nuisance parameters.

Next, we proceed in setting limits, and estimating significances. The modified frequentist criterion~\cite{Junk1999,bib-cls} is used to set 
upper  limits on the signal cross sections, following the prescription described in  Refs.~\cite{ATLAS:1379837,Cowan:2010js}, using the asymptotic 
approximation of the test statistic. In cases where the small number of events would make this approximation less accurate, the limit and significance estimation with respect to the background-only hypothesis is performed using test-statistic distributions derived from pseudo-experiments.

\begin{figure}[ht]
    \centering
    \includegraphics[width=0.90\textwidth]{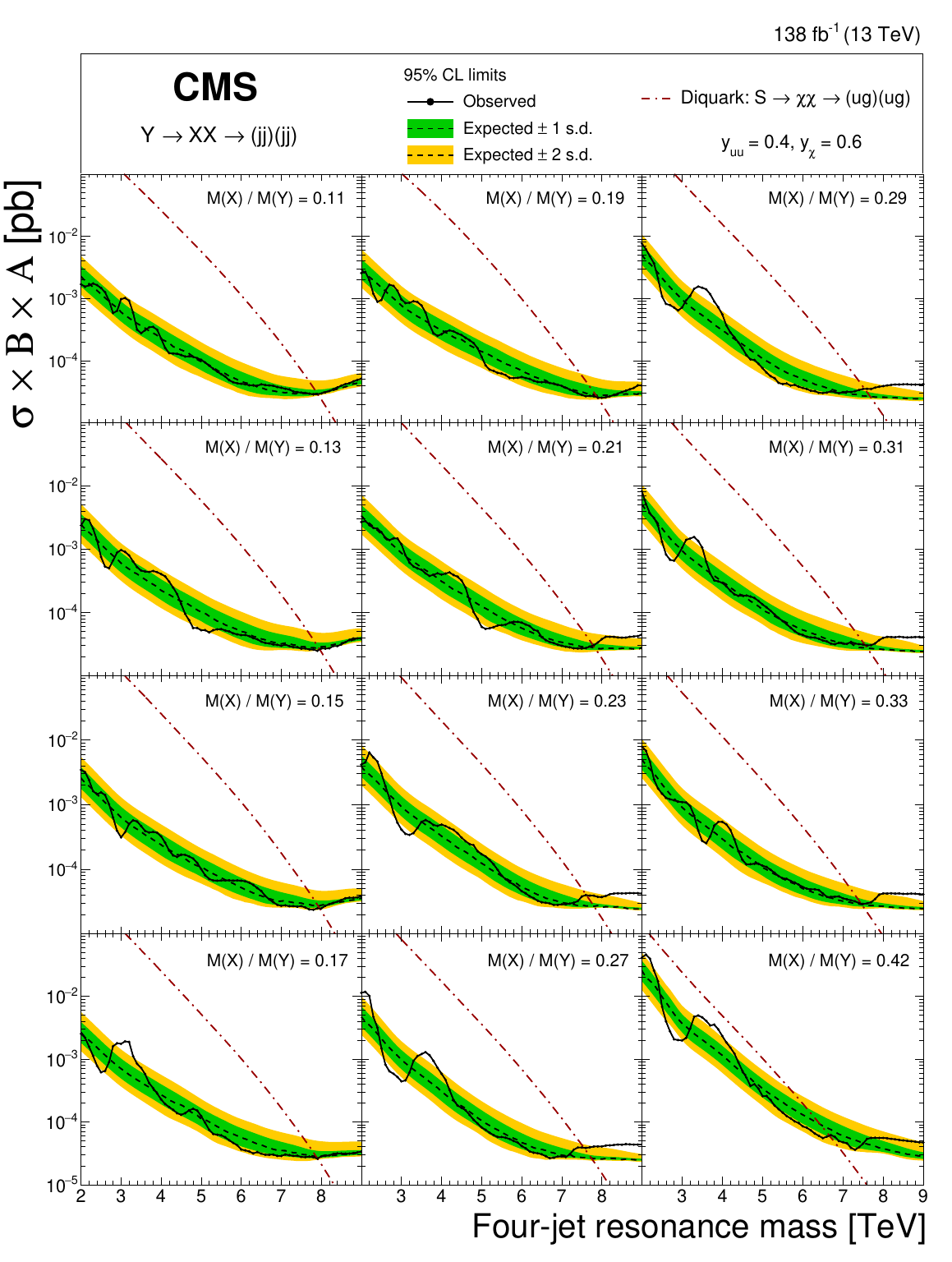}
    \caption{The observed 95\% \CL upper limits (black lines with points) on the product of the cross section, branching fraction, and acceptance for resonant
    production of  paired dijet resonances  decaying to a quark-gluon pair, with the values of $\alpha_{\mathrm{true}} = M(\PX)/M(\PY) $ shown.
    The corresponding expected limits (dashed lines) and their variations at the 1 and 2 standard deviation levels (shaded bands) are also shown.
    Limits are compared to predictions for a scalar diquark~\cite{Robertpaper} (dot-dashed line) with couplings to pairs of up quarks, $y_{uu}$= 0.4, and to pairs of vector-like quarks, $y_{\chi}$= 0.6.}
     \label{fig_limits_all}
  \end{figure}

\begin{figure}[ht]
    \centering
    \includegraphics[width=0.9\textwidth]{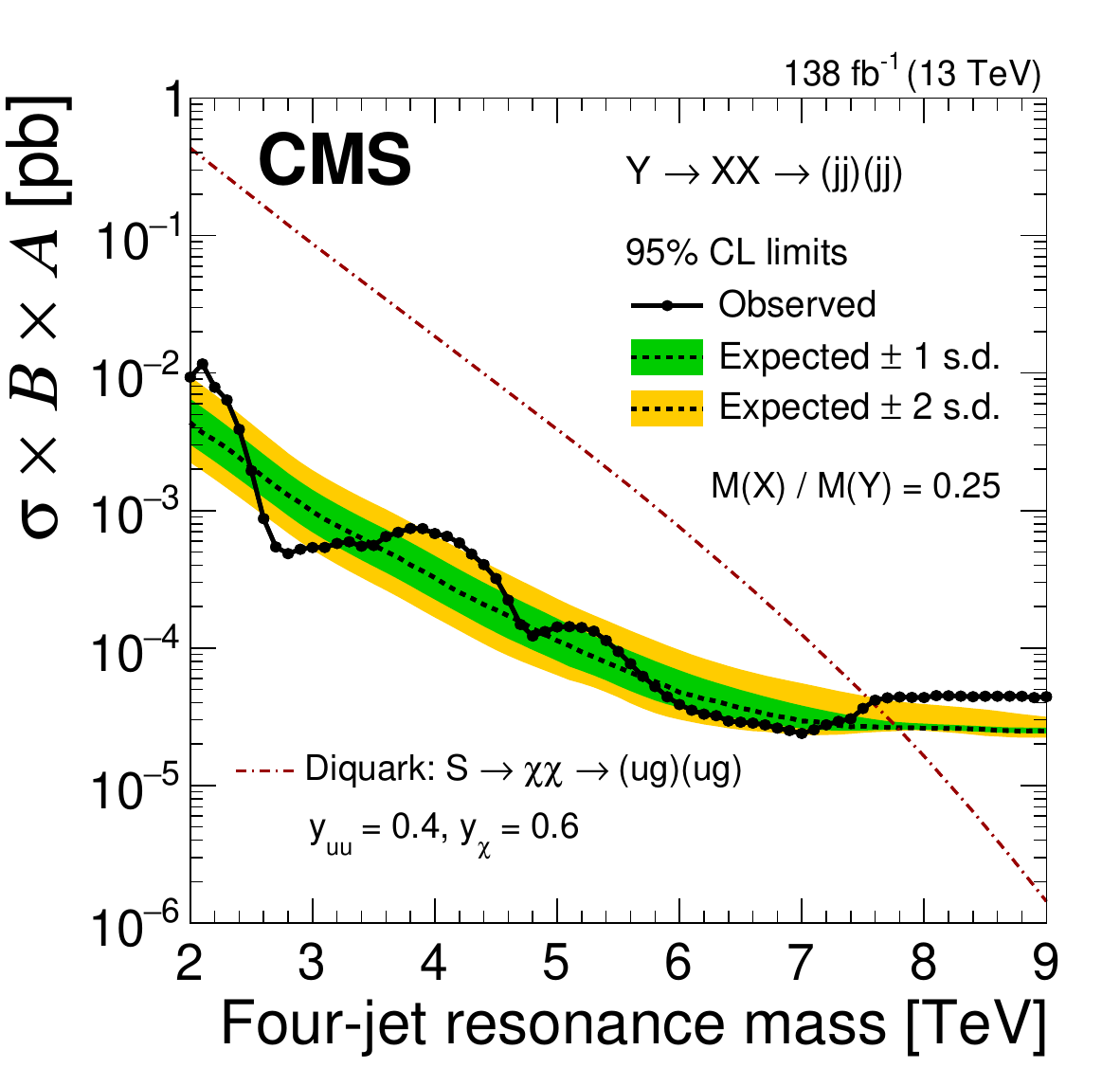}
    \caption{The observed 95\% \CL upper limits (black lines with points) on the product of the cross section, branching fraction, and acceptance for resonant
    production of  paired dijet resonances  decaying to a quark-gluon pair, with $\alpha_{\mathrm{true}}=M(\PX)/M(\PY)=0.25$.
    The corresponding expected limits (dashed lines) and their variations at the 1 and 2 standard deviation levels (shaded bands) are also shown.
   Limits are compared to predictions for a scalar diquark~\cite{Robertpaper} (dot-dashed line) with couplings to pairs of up quarks, $y_{uu}$= 0.4, and to pairs of vector-like quarks, $y_{\chi}$= 0.6.}
     \label{fig_limits}
  \end{figure}

\begin{figure}[ht]
    \centering
    \includegraphics[width=0.9\textwidth]{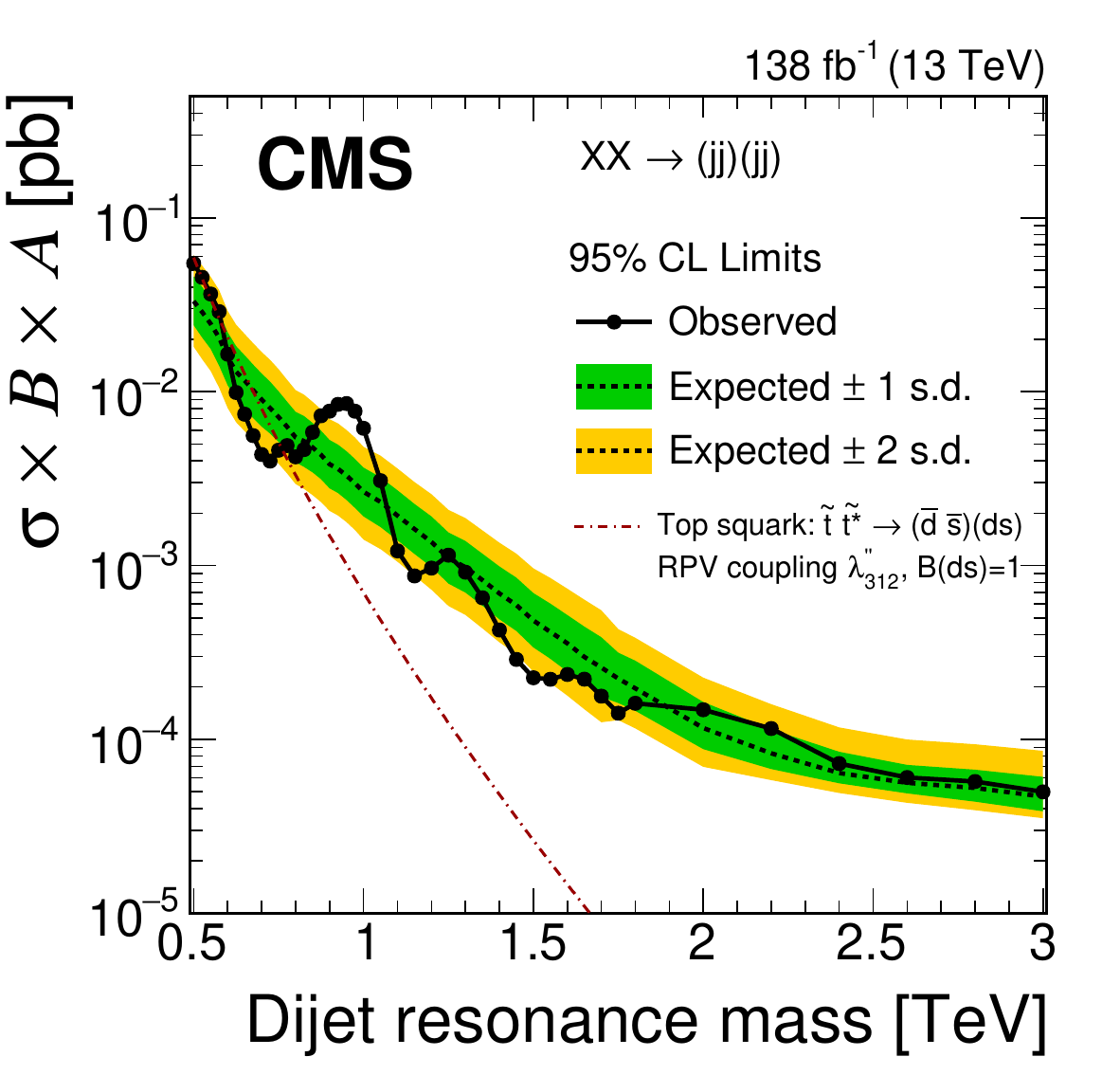}
    \caption{The observed 95\% \CL upper limits (black lines with points) on the product of the cross section, branching fraction, and acceptance for the nonresonant production of top squark pairs in the RPV SUSY decay scenario. The corresponding expected limits (dashed lines) and their variations at the 1 and 2 standard deviation levels (shaded bands) are also shown.
    Limits are compared to the top squark model cross section (dot dashed lines).} 
    \label{fig:LimNonRes}
 \end{figure}

Figures~\ref{fig_limits_all}--\ref{fig:LimNonRes} show the model-independent observed and expected upper limits at 95\% 
confidence level (\CL) on $\sigma\ \mathcal{B} A$, which is the product of the cross  section ($\sigma$), the  branching fraction ($\mathcal{B}$), and the acceptance ($A$) for the kinematic requirements. Figure~\ref{fig_limits_all} shows the limits on resonant production for twelve values of $\alpha_{\mathrm{true}}$, chosen at the center of each bin of $\alpha$, and Fig.~\ref{fig_limits} shows the limit for $\alpha_{\mathrm{true}}=0.25$, for which the maximum significance is observed.
All upper limits presented can be compared to the predictions of $\sigma\ \mathcal{B} A$ to determine 
mass limits on new particles.  In Figs.~\ref{fig_limits_all} and~\ref{fig_limits} we compare these limits from the resonant search, to 
the predicted cross section times branching fraction for the 
scalar diquark model, multiplied  by the acceptance in Fig.~\ref{fig:acc_SuuMass}. For the prediction we use a next-to-leading order (NLO) calculation with NNPDF NLO 3.1 PDFs, 
discussed in Ref.~\cite{Robertpaper}, for diquark masses greater than 6\TeV. For lower diquark masses the NLO calculation is not 
available, so we use a  \MGvATNLO calculation 
multiplied by a $K$ factor of 1.47, which is the ratio of the NLO and \MADGRAPH cross sections at 6\TeV. In Fig.~\ref{fig:LimNonRes}, the top squark
signal cross section is calculated from a simplified topology by the LHC
SUSY cross section working group at NLO including next-to-leading logarithms~\cite{SUSYxsec,nllfast31}.

For a given model, new particles are excluded at 95\% \CL in mass regions where the theoretical prediction lies at or above the 
observed upper limit for the appropriate final state in Figs.~\ref{fig_limits_all}--\ref{fig:LimNonRes}. From Fig.~\ref{fig_limits_all}, the 95\% \CL lower bound on the mass of a benchmark model of a 
diquark, for $\alpha_{\mathrm{true}}$ values between 0.11 and 0.42, varies between 7.9 and 6.7\TeV (observed), and between 7.9 and 6.4\TeV  (expected), respectively. From Fig.~\ref{fig_limits}, for 
$\alpha_{\mathrm{true}} = 0.25$ we exclude at 95\% \CL diquark masses less than 7.6\TeV (observed) and 7.8\TeV (expected).  These limits are for a benchmark  choice of coupling values of the diquark to 
pairs of up quarks, $y_{uu} = 0.4$, and to pairs of vector-like quarks, $y_{\chi} = 0.6$. Other choices of coupling values can yield different values for the diquark cross sections and the mass limits~\cite{Robertpaper}. From Fig.~\ref{fig:LimNonRes}, we exclude at 95\% \CL top squarks with masses between  
0.50 and 0.52\TeV, and between 0.58 and 0.77\TeV (observed), and less than 0.67\TeV (expected).

The observed limits in the resonant and the nonresonant search are highly correlated, because they use the same events to search for  
localized excesses (bumps), though they search for different signals. Figure~\ref{fig:LimNonRes} shows that in the 
nonresonant search, at a dijet resonance mass of roughly 0.95\TeV, the observed limit is more than 2 standard deviations (s.d.) above the expected limit. This is the direct result of the pulls near an 
average dijet mass of 1\TeV for $0.25 < \alpha <0.35$ in Fig.~\ref{fig:nonresslices} which resulted in a low $p$-value for the background fit, as previously discussed. There is a correlated effect 
in the resonant search. In Fig.~\ref{fig_limits_all} we see that the  observed limit is also more than 2 s.d. above the expected limit for a four-jet resonance mass between 3 and 
4\TeV for $\alpha_{\mathrm{true}} =$ 0.27, 0.29, and 0.31.

\clearpage

\begin{figure}[ht]
\centering
\includegraphics[width=0.9\textwidth]{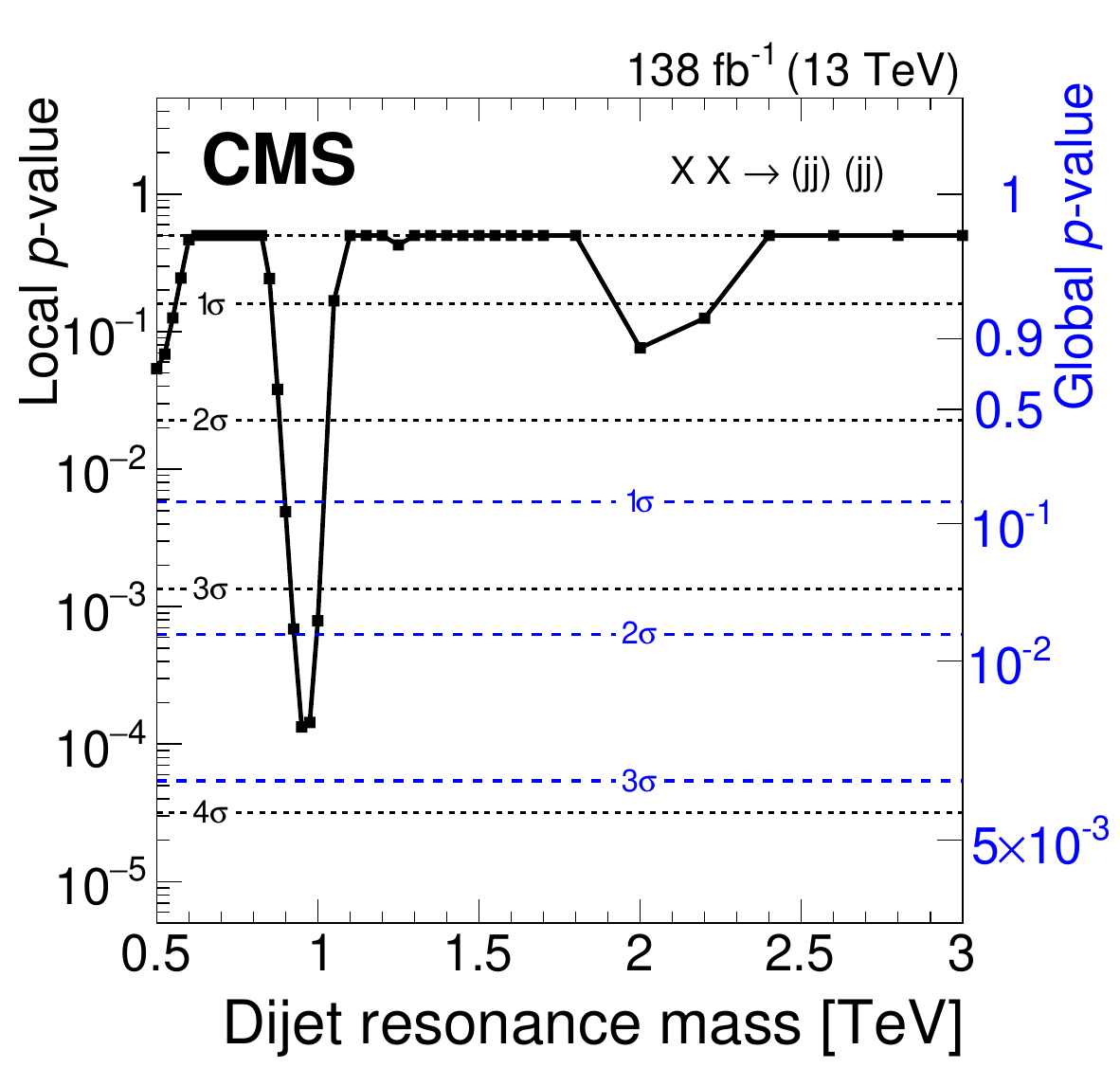}
\caption{Observed $p$-values (points) for nonresonant production of a paired dijet resonance, \PX. The vertical scales indicate the local $p$-value (left axis) and the global $p$-value (right axis in blue) for a signal over all $\alpha$ bins. Also shown are corresponding levels of local significance (dashed) and global significance (blue dashed) in units of standard deviation ($\sigma$).}
\label{fig_significance_nonres}
\end{figure}

For the nonresonant search, Fig.~\ref{fig_significance_nonres} shows the local $p$-value as a function of dijet resonance 
mass. The most significant  signal hypothesis occurs at a dijet resonance mass of 0.95\TeV, for which the local 
significance is 3.6 s.d. and the global significance is 2.5 s.d.. Here the global 
significance takes into account the look-elsewhere effect (LEE)~\cite{GrossVitells}, using pseudo-experiments to calculate the probability of 
the background hypothesis to produce a signal-like effect with at least the observed local significance, anywhere within 
the sensitive range of dijet resonance mass.

\begin{figure}[ht]
\centering
\includegraphics[width=0.9\textwidth]{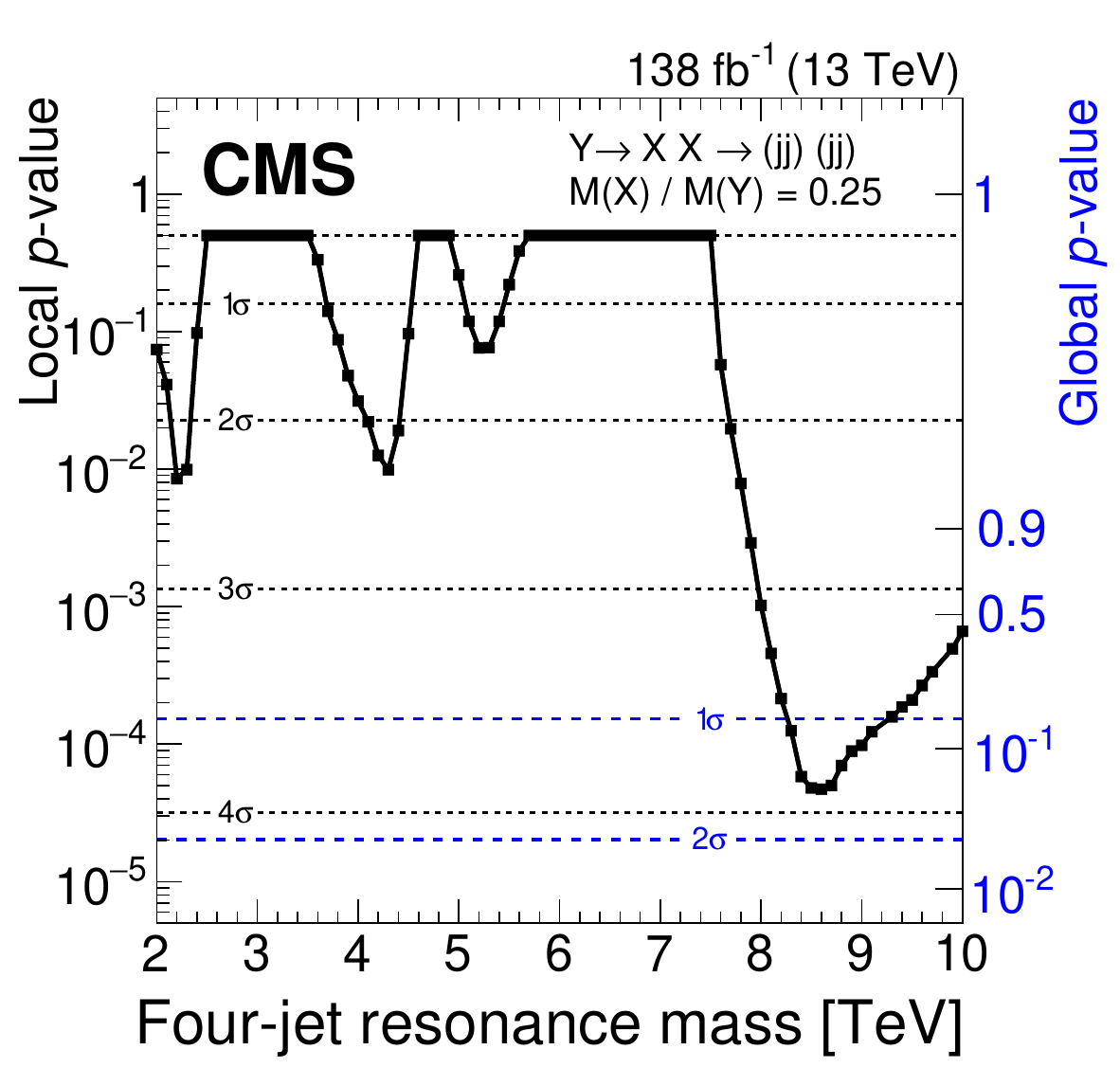}
\caption{Observed $p$-values (points) for a four-jet resonance, Y, decaying to a pair of dijet resonances, \PX, with $\alpha_\textrm{true} = M(\PX)/M(\PY) = 0.25$. The vertical scales indicate the local $p$-value (left axis) and the global $p$-value (right axis in blue) for a signal over all $\alpha$ bins. Also shown are corresponding levels of local significance (dashed) and global significance (blue dashed) in units of standard deviation ($\sigma$).}
  \label{fig_significance}
  \end{figure}

For the resonant search, Fig.~\ref{fig_significance} shows the local $p$-value as a function of four-jet resonance mass 
with $\alpha_{\mathrm{true}}=0.25$. The most significant signal hypothesis occurs at a four-jet resonance mass 
of 8.6\TeV and a dijet resonance mass of 2.1\TeV, for which the local and global significance are 3.9 s.d. and 1.6 s.d., respectively. 
Here again the global significance takes into account the LEE, using 
pseudo-experiments as described for the nonresonant search, but the LEE for the resonant search considers a 
significantly larger number of possible signals because of the 2D sensitive range of four-jet resonance mass and 
$\alpha_{\mathrm{true}}$. At a four-jet resonance mass of 8.6\TeV the best fit value of $\sigma\ \mathcal{B} A$ is ($1.56\substack{+1.40 \\ 
-0.87}$) $10^{-5}$\unit{pb}, which corresponds to $2.1\substack{+2.0\\ -1.2}$ events. The second largest local significance is 
2.9 s.d., for a four-jet resonance mass of $\sim$3\TeV and $\alpha_{\mathrm{true}}=0.17$. We note that the LEE assumes equal significance to an effect for any value of resonance mass, 
and the resonance width being smaller in absolute units at low masses yields higher probability for an effect within the low mass region.

\clearpage

\section{Candidate Events}
\label{sec:Events}

\begin{figure}[ht]
    \centering
    \includegraphics[width=0.9\textwidth]{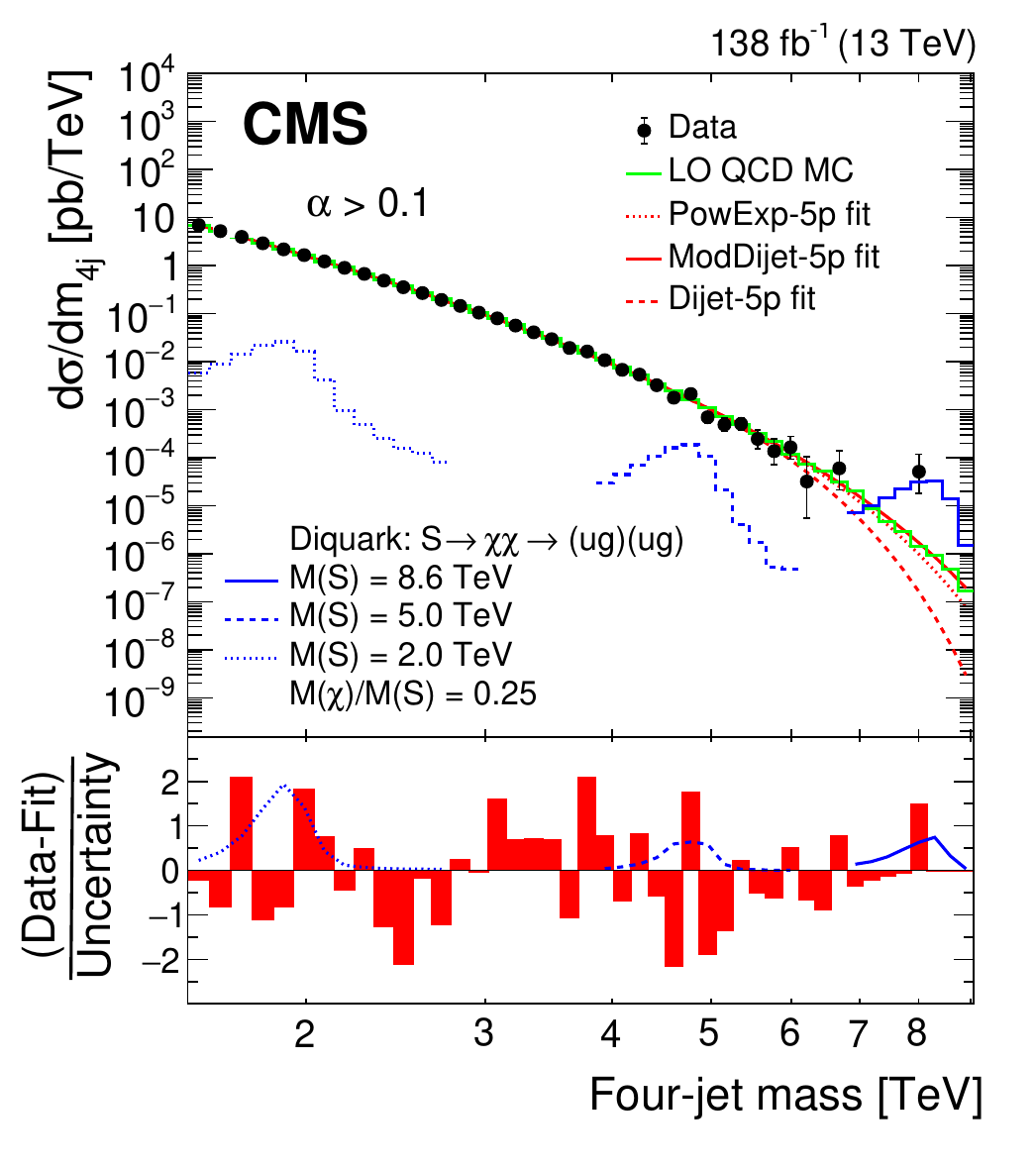}
    \caption{The four-jet mass distributions of the data (points) for all $\alpha$ bins together. The data are compared with an LO QCD prediction of 
     jet production from \PYTHIA8 (green histogram) and fitted with three functions: a power-law times an exponential (red 
     dotted), the dijet function (red dashed), and  the modified dijet function (red solid), each function has five free parameters. 
     Examples of predicted diquark signals, with cross sections equal to the observed upper limits at 
     95\% \CL, are shown for resonance masses of 2.0 (blue  dotted), 5.0 (blue dashed), and 8.6\TeV (blue solid). The lower panel shows the pulls from the 
     fit of the modified dijet function to the data.}
  \label{fig_inclusive}
 \end{figure}

The high local significance, and corresponding global significance, of the effect observed at 8.6\TeV is predominantly caused by the two events which we previously identified, and we discuss them in more detail here and in the next two paragraphs. They are in  
Fig.~\ref{fig:M4J_vs_alpha}, with a four-jet mass of roughly 8\TeV and an average dijet mass of roughly 2\TeV, and in Fig.~\ref{fig:M4J_vs_alpha_res} with the same four-jet mass.
To better 
visualize the comparison of this small number of events to the distributions of the rest of the events, the background, and the signal hypothesis, 
we show in Fig.~\ref{fig_inclusive} the inclusive 
four-jet mass distribution  corresponding to the sum of all  thirteen $\alpha$ bins. This sums the two events which appear in Fig.~\ref{fig_dijet2} in two separate plots corresponding to $\alpha$ 
bins with $0.22 < \alpha < 0.24$ and $0.26 < \alpha < 0.28$. In this case, 
the background-only fits are performed using the same functional forms, but with five parameters due to the larger event samples. These are the modified dijet function (ModDijet-5p) given by $\rd\sigma/\rd{m}= p_0(1-x^{1/3})^{p_1}/(x^{p_2 + p_3 \log{x} + p_4 \log^2{x}})$, the dijet function (Dijet-5p) given by $\rd\sigma/\rd{m}= p_0(1-x)^{p_1}/(x^{p_2 +p_3 \log{x} + p_4 \log^2{x}})$,
and the power-law times exponential function (PowExp-5p) given by $\rd\sigma/\rd{m}=p_0\re^{-p_1 x -p_2 x^2- p_3 x^3}/x^{p_4}$, where $x=m/\sqrt{s}$, with $m$ being the four-jet mass.

\begin{figure}[ht]
\centering
\includegraphics[width=0.9\textwidth]{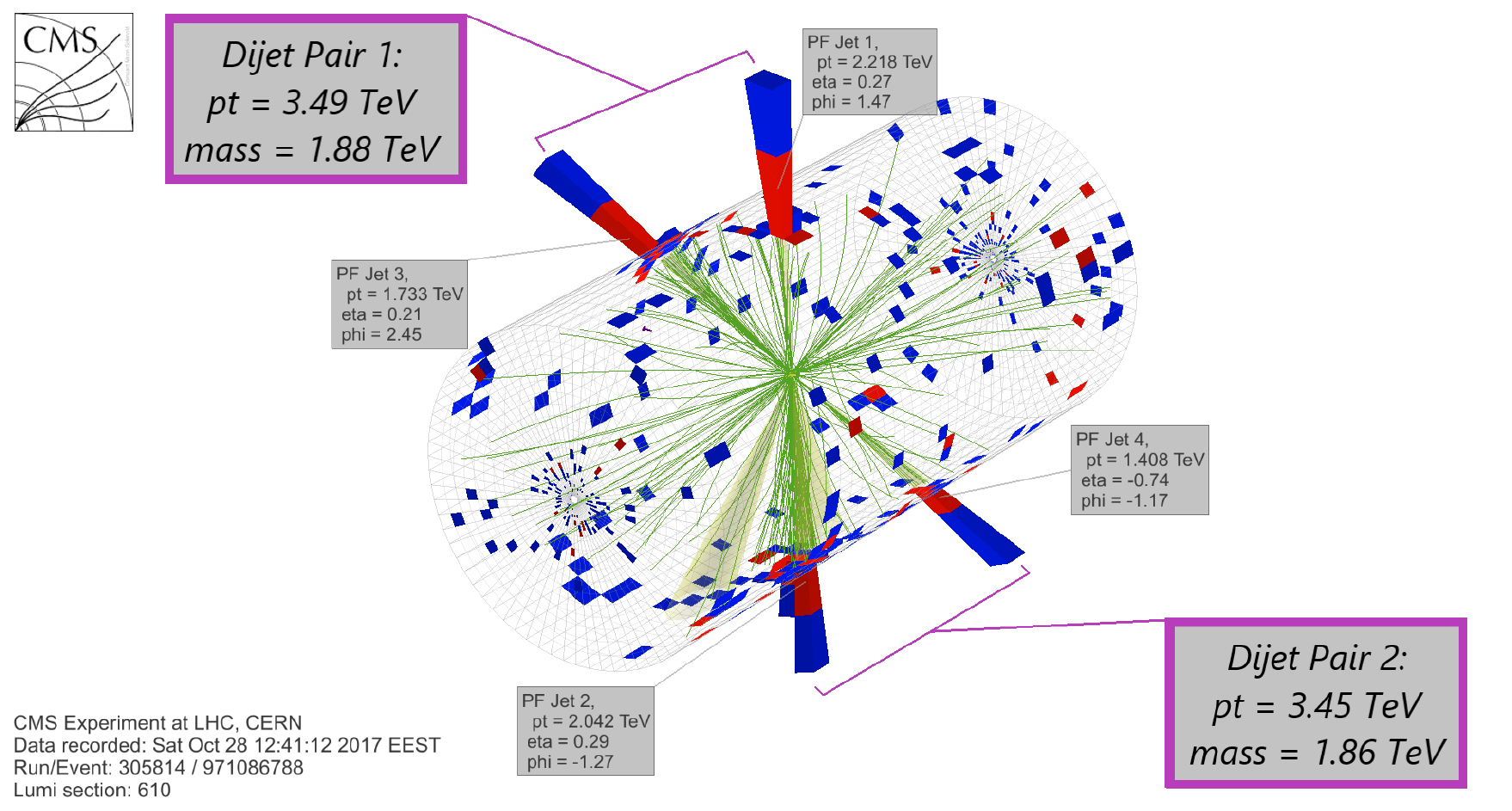}
 \caption{Three--dimensional display of the event with the highest four-jet mass of 8.0\TeV. The display shows the energy deposited in the 
electromagnetic (red) and hadronic 
(blue) calorimeters and the
reconstructed tracks of charged particles (green). The grouping of the four observed jets into two dijet pairs (purple box) is discussed in the text.}
\label{fig:eventdisplay1}
\end{figure}

The first event, shown in Fig.~\ref{fig:eventdisplay1}, was collected during 2017, and 
has a four-jet mass of 8.0\TeV, while the invariant mass of both dijet pairs is 1.9\TeV. 
This is the event found in the CMS search in the dijet final state~\cite{Sirunyan:2019vgj} which provided some of the motivation for this search.
The \pt of the first dijet pair,  
composed of jets labeled 1 and 3, is 3.5\TeV, and the second pair, consisting of jets 2 and 4, also has a \pt of 3.5\TeV. The 
two dijet pairs are back-to-back in azimuthal angle ($\abs{\Delta\phi}=3.1$), nearby in pseudorapidity 
($\abs{\Delta\eta}=0.4$) and have a small mass asymmetry of 0.005. The two jets in each pair have a small distance with each other in the $\eta$-$\phi$ plane ($\Delta R_{1}=1.0,\ \Delta 
R_{2}=1.0$), and the \pt, $\eta$, and $\phi$ of each jet are shown in Fig.~\ref{fig:eventdisplay1}. 
Since the previous search~\cite{Sirunyan:2019vgj}, a recalibration of 
the jet energy scale caused the \pt and mass values presented for this event to increase by roughly 1\%.

\begin{figure}[ht]
\centering
\includegraphics[width=0.9\textwidth]{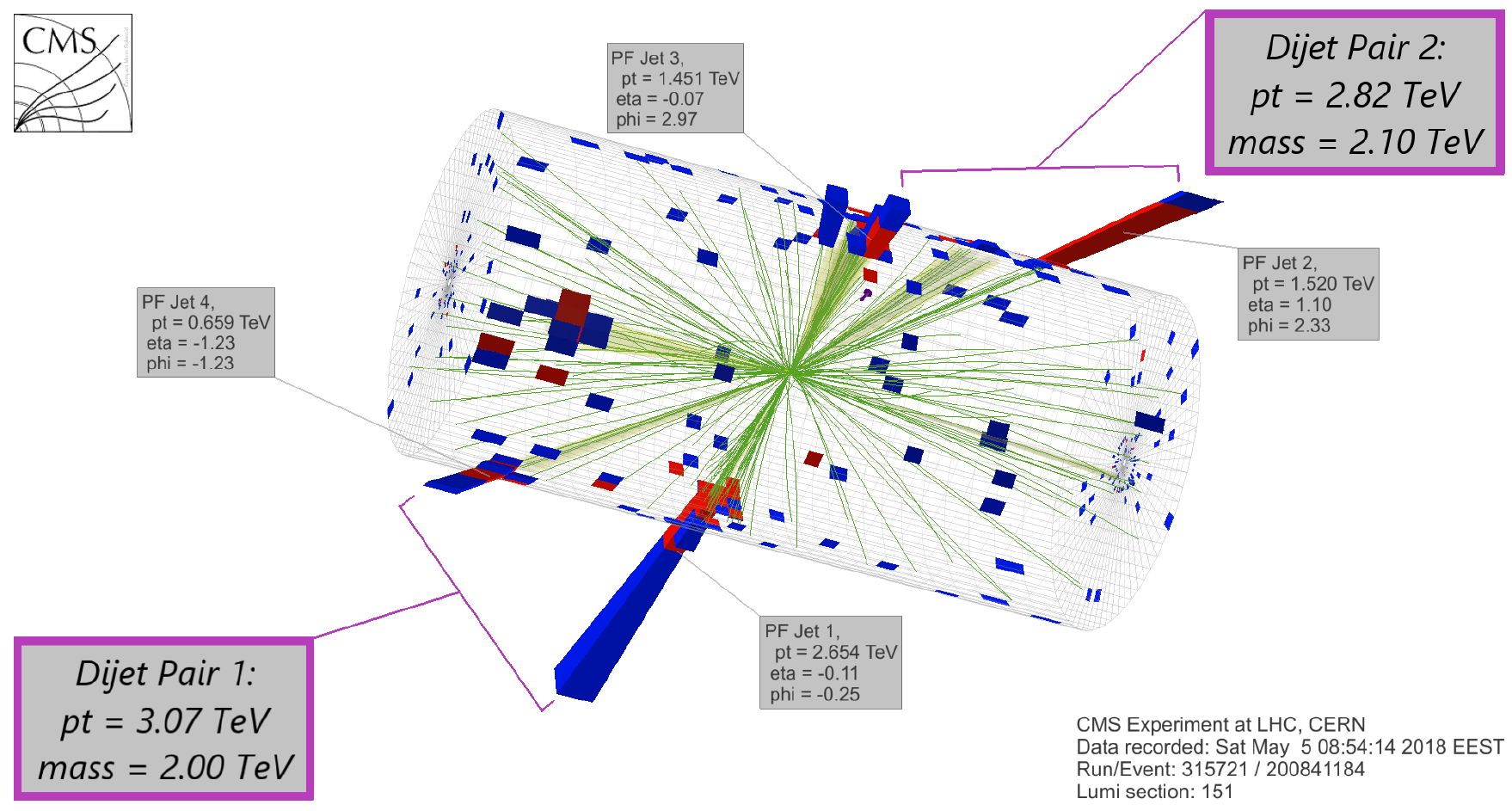}
 \caption{Three--dimensional display of the event with the second-highest four-jet mass of 7.9\TeV. The display shows the energy deposited in the 
electromagnetic (red) and hadronic (blue) calorimeters and the
reconstructed tracks of charged particles (green). The grouping of the four observed jets into two dijet pairs (purple box) is discussed in the text.}
\label{fig:eventdisplay2}
\end{figure}

The second event, shown in Fig.~\ref{fig:eventdisplay2}, was collected during 2018, and has a four-jet mass of 7.9\TeV, while the invariant 
mass of the first dijet pair that includes jets 1 and 4 is 2.0\TeV, and the mass of the second pair from jets 2 and 3 is 2.1\TeV. The two dijet pairs are back-to-back in azimuthal angle 
($\abs{\Delta\phi}=3.1$), nearby in pseudorapidity ($\abs{\Delta\eta}=1.06$), and have a mass 
asymmetry of 0.02. The $\Delta R_{1}$ and $\Delta R_{2}$ values between the two jets in the two dijet pairs are 1.5 and 1.3. 
Again, the \pt, $\eta$, and $\phi$ of each jet can be seen in Fig.~\ref{fig:eventdisplay2}.

\clearpage

\section{Summary}

A search for paired dijet resonances has been presented in final states with at least four jets, probing dijet masses above 0.35\TeV and
four-jet masses above 1.6\TeV. Data from proton-proton collisions at $\sqrt{s}=13\TeV$ were used in this search, collected by the CMS 
experiment at the LHC, corresponding to an integrated luminosity of \RunLumi. 

For the first time, a search for resonant production of pairs of dijet resonances, where a massive intermediate state leads to a four-jet resonance in the final state, has been performed.
A search for pairs of equal mass dijet resonances produced by a nonresonant process has also been conducted.
Empirical functions that model the background, and simulated shapes of resonance signals, are fit to 
the observed four-jet and dijet mass distributions. There are two events in the tails of the distributions, with a four-jet mass 
of 8\TeV and an average dijet mass of 2\TeV. They result in a local 
significance of 3.9 standard deviations and a global significance of 1.6 standard deviations, when interpreted as a signal of a 
four-jet resonance with mass 8.6\TeV and a dijet resonance with mass 
2.15\TeV. More data are needed to establish if these events are the first hints of a signal, and LHC data with the highest 
possible collision energy would be very effective.

Model-independent upper limits at 95\% confidence level (\CL) are presented on the production cross section times branching fraction and 
acceptance. For resonant production,
the limits are presented as a function of the four-jet resonance mass between 2 and 9\TeV, for all accessible values of the ratio of 
the dijet to four-jet resonance
masses. For nonresonant production, the limits are presented as
functions of dijet resonance mass between 0.5 and 3.0\TeV.

Limits from the resonant search are compared to a model~\cite{Robertpaper} of diquarks, which decay to pairs of vector-like 
quarks, which in-turn decay to a quark and a gluon. Mass 
limits for all accessible values of the ratio of the vector-like quark to diquark masses, for a benchmark scenario where the diquark couplings to pairs of up quarks is 0.4, and the diquark couplings 
to pairs of  vector-like quarks is 0.6 are presented. In particular, when the ratio of the vector-like quark to diquark mass is chosen to be 0.25, motivated by the two observed events discussed above,  diquarks with a mass less than 7.6\TeV (observed) and 7.8\TeV (expected) are excluded at 95\% \CL.

Limits from the nonresonant search are compared to a model~\cite{Barbier:2004ez} of $R$-parity-violating supersymmetry, with pair produced top squarks 
decaying to down and strange quarks. 
Top squarks with masses between 0.50\TeV to 0.77\TeV are excluded at 95\% \CL,
with the exception of a small interval between 0.52 and 0.58\TeV, compared with an exclusion of less than 0.67\TeV (expected), significantly extending the previous observed limit of 0.52\TeV~\cite{Sirunyan:2018rlj}. In this case, the most significant 
signal hypothesis occurs at an average dijet mass of 0.95\TeV, for which the local and global significances are 3.6 and 2.5 standard deviations, respectively.

\begin{acknowledgments}
We thank Bogdan Dobrescu and Josh Isaacson for supplying the code and cross section for the diquark model and for their work developing a benchmark model specifically for this search.

We congratulate our colleagues in the CERN accelerator departments for the excellent performance of the LHC and thank the technical and administrative staffs at CERN and at other CMS institutes for their contributions to the success of the CMS effort. In addition, we gratefully acknowledge the computing centers and personnel of the Worldwide LHC Computing Grid and other centers for delivering so effectively the computing infrastructure essential to our analyses. Finally, we acknowledge the enduring support for the construction and operation of the LHC, the CMS detector, and the supporting computing infrastructure provided by the following funding agencies: BMBWF and FWF (Austria); FNRS and FWO (Belgium); CNPq, CAPES, FAPERJ, FAPERGS, and FAPESP (Brazil); MES and BNSF (Bulgaria); CERN; CAS, MoST, and NSFC (China); MINCIENCIAS (Colombia); MSES and CSF (Croatia); RIF (Cyprus); SENESCYT (Ecuador); MoER, ERC PUT and ERDF (Estonia); Academy of Finland, MEC, and HIP (Finland); CEA and CNRS/IN2P3 (France); BMBF, DFG, and HGF (Germany); GSRI (Greece); NKFIH (Hungary); DAE and DST (India); IPM (Iran); SFI (Ireland); INFN (Italy); MSIP and NRF (Republic of Korea); MES (Latvia); LAS (Lithuania); MOE and UM (Malaysia); BUAP, CINVESTAV, CONACYT, LNS, SEP, and UASLP-FAI (Mexico); MOS (Montenegro); MBIE (New Zealand); PAEC (Pakistan); MES and NSC (Poland); FCT (Portugal); MESTD (Serbia); MCIN/AEI and PCTI (Spain); MOSTR (Sri Lanka); Swiss Funding Agencies (Switzerland); MST (Taipei); MHESI and NSTDA (Thailand); TUBITAK and TENMAK (Turkey); NASU (Ukraine); STFC (United Kingdom); DOE and NSF (USA).

\hyphenation{Rachada-pisek} Individuals have received support from the Marie-Curie program and the European Research Council and Horizon 2020 Grant, contract Nos.\ 675440, 724704, 752730, 758316, 765710, 824093, 884104, and COST Action CA16108 (European Union); the Leventis Foundation; the Alfred P.\ Sloan Foundation; the Alexander von Humboldt Foundation; the Belgian Federal Science Policy Office; the Fonds pour la Formation \`a la Recherche dans l'Industrie et dans l'Agriculture (FRIA-Belgium); the Agentschap voor Innovatie door Wetenschap en Technologie (IWT-Belgium); the F.R.S.-FNRS and FWO (Belgium) under the ``Excellence of Science -- EOS" -- be.h project n.\ 30820817; the Beijing Municipal Science \& Technology Commission, No. Z191100007219010; the Ministry of Education, Youth and Sports (MEYS) of the Czech Republic; the Hellenic Foundation for Research and Innovation (HFRI), Project Number 2288 (Greece); the Deutsche Forschungsgemeinschaft (DFG), under Germany's Excellence Strategy -- EXC 2121 ``Quantum Universe" -- 390833306, and under project number 400140256 - GRK2497; the Hungarian Academy of Sciences, the New National Excellence Program - \'UNKP, the NKFIH research grants K 124845, K 124850, K 128713, K 128786, K 129058, K 131991, K 133046, K 138136, K 143460, K 143477, 2020-2.2.1-ED-2021-00181, and TKP2021-NKTA-64 (Hungary); the Council of Science and Industrial Research, India; the Latvian Council of Science; the Ministry of Education and Science, project no. 2022/WK/14, and the National Science Center, contracts Opus 2021/41/B/ST2/01369 and 2021/43/B/ST2/01552 (Poland); the Funda\c{c}\~ao para a Ci\^encia e a Tecnologia, grant CEECIND/01334/2018 (Portugal); the National Priorities Research Program by Qatar National Research Fund; MCIN/AEI/10.13039/501100011033, ERDF ``a way of making Europe", and the Programa Estatal de Fomento de la Investigaci{\'o}n Cient{\'i}fica y T{\'e}cnica de Excelencia Mar\'{\i}a de Maeztu, grant MDM-2017-0765 and Programa Severo Ochoa del Principado de Asturias (Spain); the Chulalongkorn Academic into Its 2nd Century Project Advancement Project, and the National Science, Research and Innovation Fund via the Program Management Unit for Human Resources \& Institutional Development, Research and Innovation, grant B05F650021 (Thailand); the Kavli Foundation; the Nvidia Corporation; the SuperMicro Corporation; the Welch Foundation, contract C-1845; and the Weston Havens Foundation (USA).
\end{acknowledgments}\section*{Data availability} Release and preservation of data used by the CMS Collaboration as the basis for publications is guided by the  \href{https://doi.org/10.7483/OPENDATA.CMS.1BNU.8V1W}{CMS data preservation, re-use and open access policy}.

\bibliography{auto_generated}

\appendix
\section{Search for resonant production of pairs of dijet resonances through broad mediators\label{app}}

We extend the analysis with a search for models with resonant production of a pair of dijet resonances via a broad mediator. This is the first time a search for a broad four-jet resonance decaying into a pair of BSM particles is performed at a hadron collider. The natural width of mediator \PY is considered here to be comparable to or larger than the experimental resolution, thus forming a broad resonance. In contrast, the dijet resonance \PX is characterized by a narrow intrinsic width to avoid large couplings, which are already excluded by dijet resonance searches for single \PX production~\cite{Sirunyan:2019vgj, Aad:2019hjw}. The event selection, jet pairing algorithm, and fitting approach are the same as those utilized for the narrow-resonance interpretation, since they remain optimal for a broad-resonance search, and potential biases are avoided in light of the significant excess of high-mass events already reported in the analysis.

This extension is further motivated by the presence of a third high-mass isolated event with \mjjjj = 5.8\TeV and \mjj = 2.0\TeV, in addition to the two events observed at \mjjjj = 8\TeV and \mjj = 2\TeV. The similar values of \mjj of these events, all around 2\TeV, and the wider spread of the observed \mjjjj, make them potential candidates for a broad four-jet resonance decaying into a pair of narrow dijet resonances.

Resonances decaying to pairs of dijet resonances arise in models of diquark particles decaying into vector-like quarks~\cite{Robertpaper, dobrescu2024tevscaleparticleslhcevents, dobrescu2019lhcprobes10tev, hxll-41ly} or into colored-triplet scalars~\cite{dobrescu2019lhcprobes10tev}, in extensions of the QCD gauge group~\cite{Robertpaper, Bai:2018jsr}, in supersymmetric scenarios with a single baryon-number and R-parity violating term~\cite{Bittar:2025rcw}, or in other theoretical frameworks~\cite{Crivellin:2022nms}.

As benchmarks, we consider two models of scalar diquark particles~\cite{Robertpaper}, where a color-sextet diquark, \PSuu or \PSdd, is produced from the annihilation of two up or down quarks. The dijet resonance is a vector-like quark, \PGc or \PGo, that decays at one loop into a gluon and an up or down quark, respectively. Here, in addition to searching for broad resonances decaying to pairs of dijet resonances, we also for the first time use and explicitly constrain the \PSdd model.

Events are generated for diquark masses between 2 and 10\TeV and for vector-like quark masses between 0.22 and 4.2\TeV. The vector-like quarks have a width-to-mass ratio smaller than $10^{-4}$, because their decays arise at one loop~\cite{dobrescu2024tevscaleparticleslhcevents}. For each \atrue signal scenario, the couplings \yuu and \ychi are chosen accordingly so as to generate diquark resonances with a total natural width of 1.5, 5, and 10\%, while keeping $\beta = \yuu/\ychi$ at its benchmark value of $2/3$. High-mass diquark signals with a width larger than approximately 10\% do not exhibit a clear peak at the resonance mass, and hence are outside the scope of this analysis. We set $\mathcal{B}(\PGc \to \PQu\Pg) = 1$, as is done in Ref.~\cite{Robertpaper}. The above discussion of couplings, widths, and our benchmark values is also true for \PSdd and $\omega$ with the symbolic replacement everywhere of $\PQu\to\PQd$ and $\PGc\to\PGo$.

\begin{figure}[hbtp]
\centering
\includegraphics[width=0.44\textwidth]{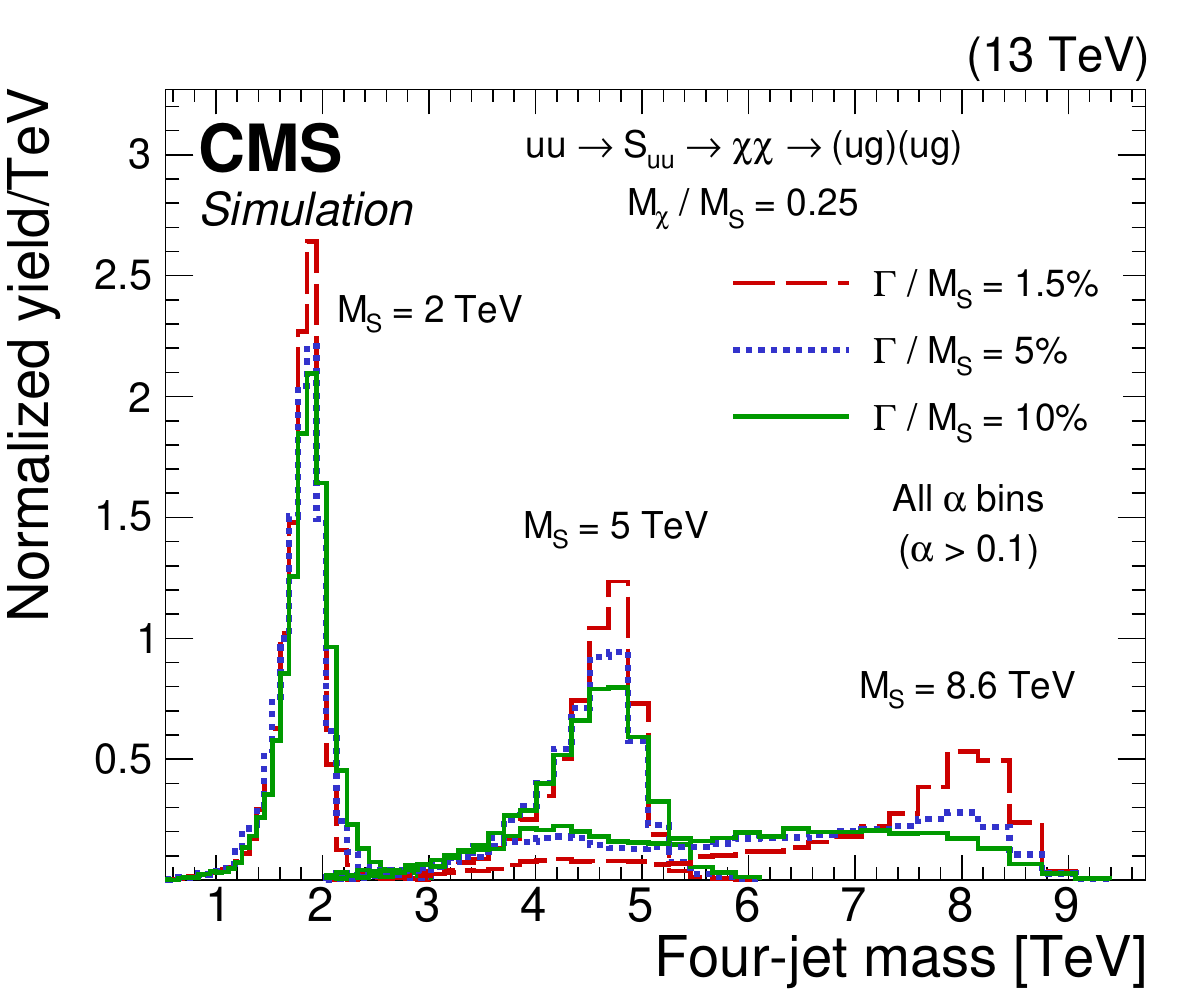} \hspace{3mm}
\includegraphics[width=0.44\textwidth]{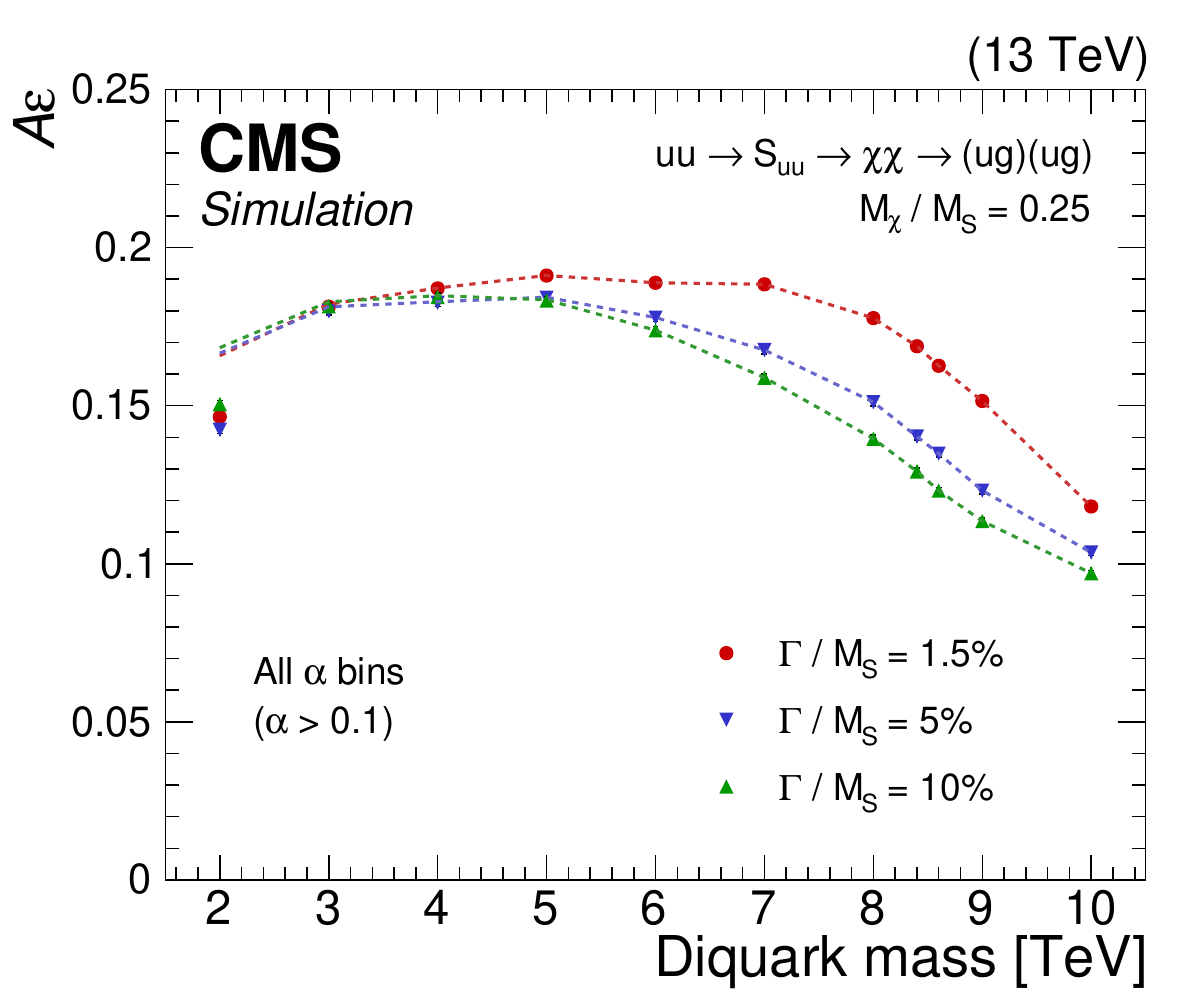}
\caption{(Left) Signal differential distributions as a function of \mjjjj for $\atrue = 0.25$, diquark masses of 2, 5, 8.6\TeV, and various widths, for all $\alpha$ bins inclusively. The integral of each distribution has been normalized to unity. (Right) The product of the acceptance ($A$) and efficiency of the mass selection ($\varepsilon$) for a resonant signal with $\atrue = 0.25$, shown with filled markers as a function of the diquark mass and for various diquark widths, for all $\alpha$ bins inclusively. The case when $\varepsilon=1$ is also shown as dashed lines.} 
\label{fig:SignalShapes_Acceptance} 
\end{figure}

Figure~\ref{fig:SignalShapes_Acceptance} (left) shows the shapes of signals with $\atrue = 0.25$ and various diquark masses, across all $\alpha$ bins. Low-mass resonances exhibit a shift toward higher \mjjjj as their natural width increases, whereas intermediate-mass resonances display a broader distribution extending to both lower and higher \mjjjj. Conversely, high-mass resonances present a long, low-mass tail, arising from the convolution of the resonance production with the parton distribution functions (PDFs). At high masses, the PDFs fall steeply with momentum fraction, increasing the probability that one or both partons have relatively low momentum, thus enhancing the low-mass tail. This effect becomes more prominent with larger natural width and larger resonance mass. Overall, a broad resonance gives an improved description of all three candidate events with $\mjj \simeq 2\TeV$.  

The products of acceptance ($A$) and efficiency of the mass selection ($\varepsilon$) are shown in Fig.~\ref{fig:SignalShapes_Acceptance} (right), as functions of diquark mass for all widths considered. The $\mjjjj > 1.6\TeV$ requirement causes $A\varepsilon$ to decrease at a four-jet resonance mass of 2\TeV. A decrease in $A\varepsilon$ is observed at the highest values of $M_{\PS}$, which is enhanced as the diquark width increases, and is found to originate from the mass asymmetry and $\Delta R$ requirements on the dijet pairs.

The background-only fits in each $\alpha$ bin were previously shown to adequately describe the data, with a combined $p$-value of 0.18 for a simultaneous fit across all thirteen $\alpha$ bins. In the present analysis, the signal-plus-background modeling was likewise examined, with goodness-of-fit $p$-values ranging from 0.18 to 0.35 across all broad resonance signal hypotheses.

The statistical treatment and systematic uncertainties for signal extraction are identical to those of the narrow four-jet resonance analysis. The background systematic uncertainties increase as the resonance width increases, an effect that is more prominent at higher resonance masses above approximately 7\TeV, where signal shapes are more affected by an increasing width, as seen in Fig.~\ref{fig:SignalShapes_Acceptance} (left). Bias studies were performed using pseudo-experiments, generated under several different signal hypotheses covering all considered widths, confirming that any bias in the signal extraction is negligible, even for the larger widths examined.

\begin{figure}[!ht]
\centering
\includegraphics[width=0.82\textwidth]{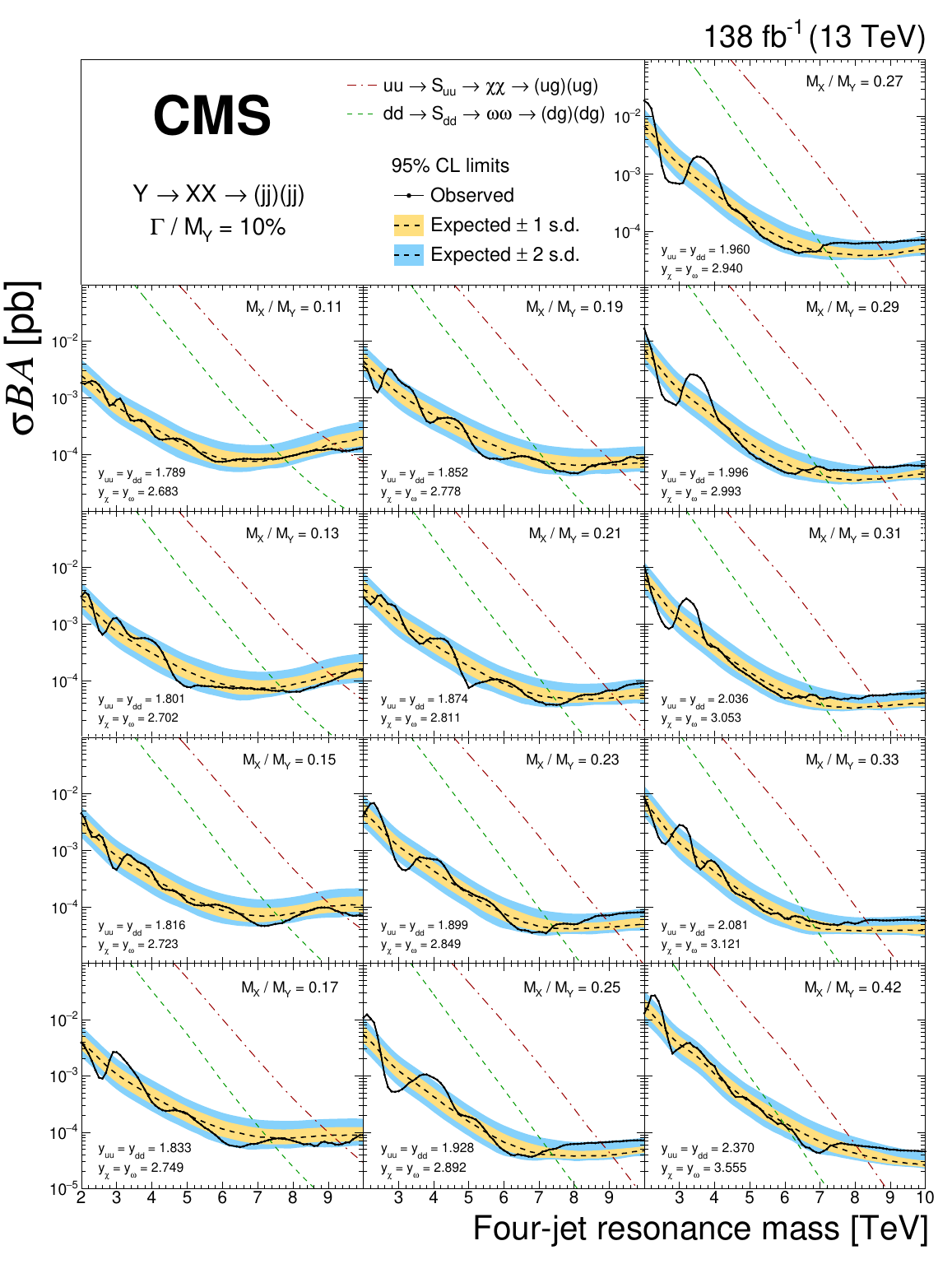} 
\caption{The observed 95\% \CL upper limits (points) on the product of the cross section, branching fraction, and acceptance for resonant production of paired dijet resonances, \PX, with width of the initial resonance \PY equal to 10\%, and values of $\atrue = M_{\PX} / M_{\PY}$ shown in each panel. The corresponding expected limits (dashed) and their variations at the 1 and 2 s.d. levels (shaded bands) are also shown. Limits are compared to predictions for scalar \PSuu (red dot-dashed lines) and \PSdd (green dashed lines) diquarks~\cite{Robertpaper} with couplings to pairs of up and down quarks, \yuu and \ydd, and to pairs of vector-like quarks, \ychi and \yomega, shown in each panel.} 
\label{fig:LimitVsMass_AllAlpha_W0p1} 
\end{figure}

The upper limits at 95\% \CL for the largest width considered in the analysis are shown in Fig.~\ref{fig:LimitVsMass_AllAlpha_W0p1}, for thirteen values of \atrue chosen at the center of each bin of $\alpha$. Exclusion mass limits for \PSuu and \PSdd signals with $\atrue = 0.25$ are presented in Table~\ref{tab:mass_limits}. For the 8.6\TeV four-jet resonance with a width of 10\%, the benchmark \PSuu diquark model is excluded, while the \PSdd diquark model remains viable and compatible with the three candidate events.

\begin{table}[th]
\centering
\topcaption{Observed and expected mass limits at 95\% \CL for \PSuu and \PSdd diquark models with $\atrue = 0.25$.}
\label{tab:mass_limits}
\begin{tabular}{lcc} 
\multirow{2}*{Width}                         & \multicolumn{2}{c}{Observed (expected) mass limit [$\TeVns$]}  \\
                                             & \PSuu scalar diquark                  & \PSdd scalar diquark   \\
\hline                               
$\Gamma / M_{\PS} = 1.5\%$  & 7.8 (8.1) &  6.5 (6.3)     \\   
$\Gamma / M_{\PS} = 5\%$    & 8.3 (8.6) &  7.0 (6.9)     \\   
$\Gamma / M_{\PS} = 10\%$   & 8.8 (9.1) &  7.3 (7.4)     \\    
\end{tabular}
\end{table}

\begin{figure}[ht!]
\centering
\includegraphics[width=0.44\textwidth]{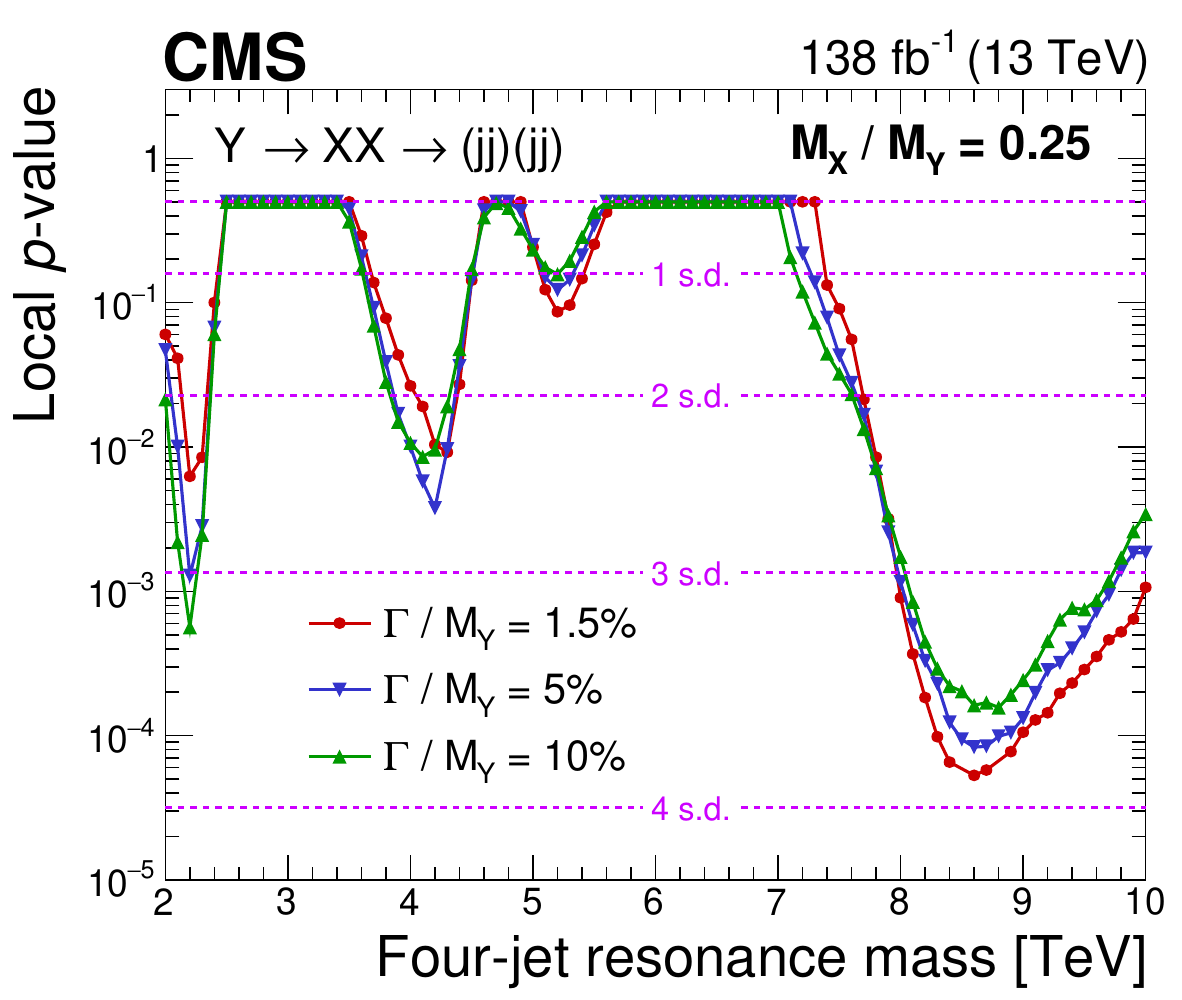} \hspace{3mm}
\includegraphics[width=0.44\textwidth]{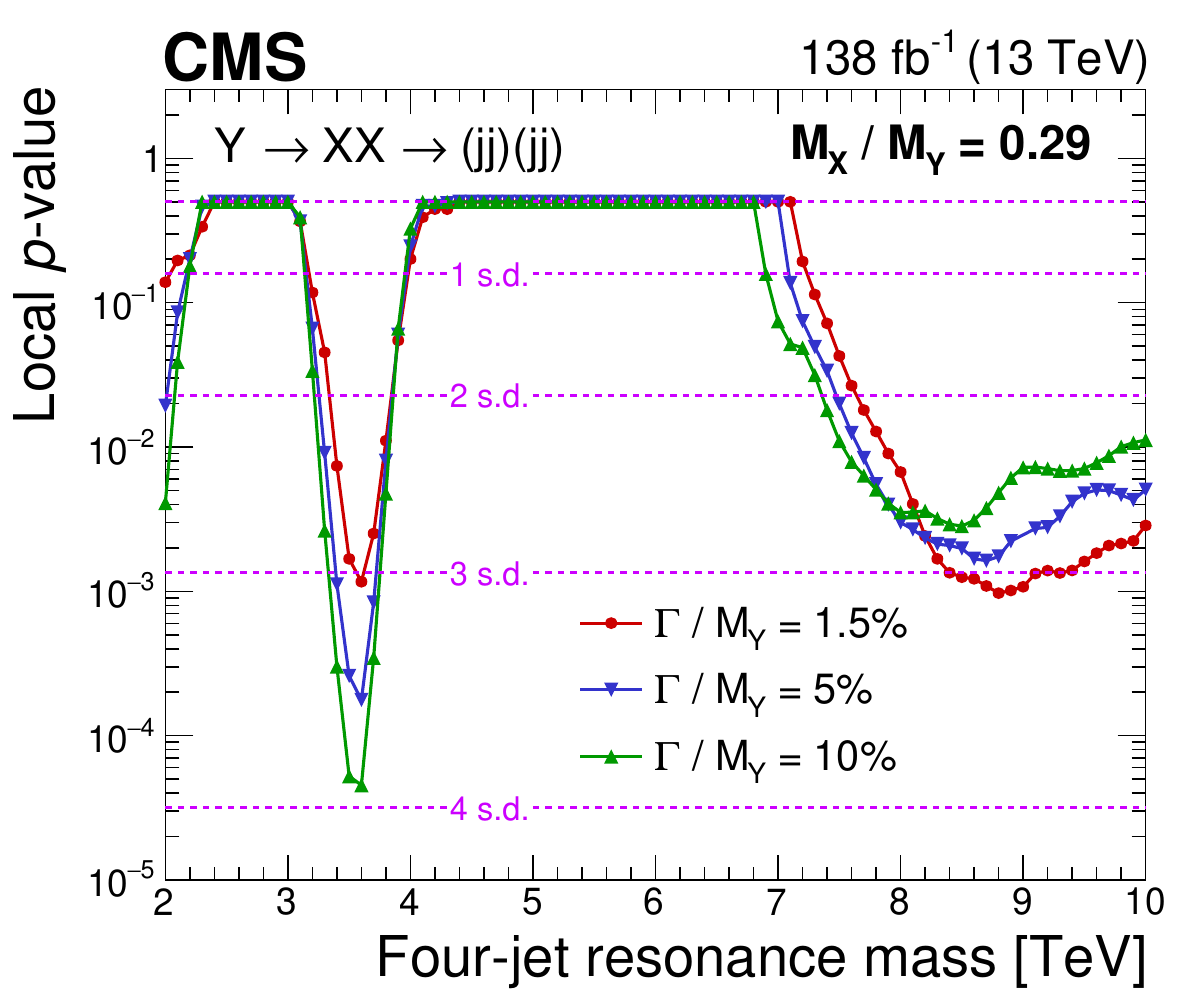}
\caption{Observed local $p$-value for a four-jet resonance, \PY, decaying to a pair of dijet resonances, \PX, with $\atrue = M_{\PX} / M_{\PY} = 0.25$ (left) and 0.29 (right), and various widths of \PY superimposed. Also shown are corresponding levels of local significance (dashed lines) in s.d.} 
\label{fig:PValueVsMassVsWidth} 
\end{figure}

Figure~\ref{fig:PValueVsMassVsWidth} (left) shows the local $p$-value of a signal, evaluated under the background-only hypothesis, as a function of the four-jet resonance mass and width for $\atrue = 0.25$. For a width of 1.5\%, the most significant signal hypothesis occurs at a four-jet resonance mass of 8.6\TeV and a dijet resonance mass of 2.15\TeV, for which the local significance is 3.9 s.d, the same as the significance observed for a narrow resonance with those mass values. As the four-jet resonance becomes broader, Fig.~\ref{fig:PValueVsMassVsWidth} demonstrates that the local significance remains high, exceeding or being equal to 3.6 s.d., even for the largest width of 10\%. Hence, the significance of an 8.6\TeV four-jet resonance is found to be relatively insensitive to the width, making a broad resonance an equally valid interpretation of the excess.

Table~\ref{tab:best_fit_r_values} summarizes the local and global significances for a four-jet resonance, along with the best fit value of $\sigma \mathcal{B} A$ and number of events, as a function of width for a four-jet resonance mass of 8.6\TeV with $\atrue = 0.25$. The table demonstrates that broader resonances, with widths of 5 and 10\%, provide a better fit for all three observed candidate events.

A second excess is observed at a four-jet resonance mass of 3.6\TeV, $\atrue = 0.29$, and 10\% width with a local and global significance of 3.9 and 2.2 s.d. respectively, as shown in Fig.~\ref{fig:PValueVsMassVsWidth} (right). This excess points to a structure in data near a dijet mass of 1\TeV, which was also reported by the search for nonresonant production of pairs of dijet resonances, and is further discussed in Refs.~\cite{dobrescu2024tevscaleparticleslhcevents, Crivellin:2022nms}. Its corresponding best fit values of $\sigma \mathcal{B} A$, number of events, and signal significance are presented in Table~\ref{tab:best_fit_r_values}. Here the significance increases with width. 

\begin{table}[!hb]
\centering
\topcaption{For each resonance width and for the two most significant resonance hypotheses found, we list the \PY and \PX resonance masses, the best fit values of signal $\sigma \mathcal{B} A$ from the signal-plus-background fit, the corresponding number of signal events, and the local and global significance.}
\label{tab:best_fit_r_values}
\renewcommand{\arraystretch}{1.3}
\begin{tabular}{lccccc} 
$\Gamma / M_{\PY}$  &   $M_{\PY}$  &  $M_{\PX}$   &    $\sigma \mathcal{B} A$  &   Events  &  Local (global) significance  \\[-1.ex] 
$[\%]$  &  $[\TeVns]$  & $[\TeVns]$   &  [pb]  &    &  [s.d.]  \\
\hline                              
1.5  & 8.6 & 2.15 &  $1.68^{+1.50}_{-0.94} \times 10^{-5}$ & $2.3^{+2.1}_{-1.3}$   &  3.9 (1.6) \\ 
5    &  "  &  "   &  $2.04^{+1.83}_{-1.15} \times 10^{-5}$ & $2.8^{+2.5}_{-1.6}$   &  3.8 (1.5) \\ 
10   &  "  &  "   &  $2.22^{+1.99}_{-1.26} \times 10^{-5}$ & $3.1^{+2.7}_{-1.7}$   &  3.6 (1.4) \\
1.5  & 3.6 &  1.0 &  $0.83^{+0.30}_{-0.32} \times 10^{-3}$ & $114^{+42}_{-44}$     &  3.0 (0.5) \\ 
5    &  "  &  "   &  $1.18^{+0.36}_{-0.33} \times 10^{-3}$ & $162^{+50}_{-46}$     &  3.6 (1.6) \\ 
10   &  "  &  "   &  $1.53^{+0.42}_{-0.40} \times 10^{-3}$ & $210^{+58}_{-54}$     &  3.9 (2.2) \\
\end{tabular}
\end{table}

We thank Bogdan Dobrescu for supplying the code and cross section for the diquark models with larger natural widths than those previously considered, and for his work developing benchmark models specifically for this search.


\cleardoublepage \section{The CMS Collaboration \label{app:collab}}\begin{sloppypar}\hyphenpenalty=5000\widowpenalty=500\clubpenalty=5000\cmsinstitute{Yerevan Physics Institute, Yerevan, Armenia}
{\tolerance=6000
A.~Tumasyan\cmsAuthorMark{1}\cmsorcid{0009-0000-0684-6742}
\par}
\cmsinstitute{Institut f\"{u}r Hochenergiephysik, Vienna, Austria}
{\tolerance=6000
W.~Adam\cmsorcid{0000-0001-9099-4341}, J.W.~Andrejkovic, T.~Bergauer\cmsorcid{0000-0002-5786-0293}, S.~Chatterjee\cmsorcid{0000-0003-2660-0349}, K.~Damanakis\cmsorcid{0000-0001-5389-2872}, M.~Dragicevic\cmsorcid{0000-0003-1967-6783}, A.~Escalante~Del~Valle\cmsorcid{0000-0002-9702-6359}, P.S.~Hussain\cmsorcid{0000-0002-4825-5278}, M.~Jeitler\cmsAuthorMark{2}\cmsorcid{0000-0002-5141-9560}, N.~Krammer\cmsorcid{0000-0002-0548-0985}, L.~Lechner\cmsorcid{0000-0002-3065-1141}, D.~Liko\cmsorcid{0000-0002-3380-473X}, I.~Mikulec\cmsorcid{0000-0003-0385-2746}, P.~Paulitsch, F.M.~Pitters, J.~Schieck\cmsAuthorMark{2}\cmsorcid{0000-0002-1058-8093}, D.~Schwarz\cmsorcid{0000-0002-3821-7331}, R.~Sch\"{o}fbeck\cmsorcid{0000-0002-2332-8784}, S.~Templ\cmsorcid{0000-0003-3137-5692}, W.~Waltenberger\cmsorcid{0000-0002-6215-7228}, C.E.~Wulz\cmsAuthorMark{2}\cmsorcid{0000-0001-9226-5812}
% ...
\par}
$^{1}$Also at Yerevan State University, Yerevan, Armenia\\
$^{2}$Also at Technische Universit\"{a}t Wien, Vienna, Austria\\
% ...
\end{sloppypar}
\end{document}